\documentclass[useAMS,usenatbib]{mn2e}

\usepackage{amsmath,natbib,graphicx}
\usepackage{url}

\newcommand{\kms}{ km s$^{-1}$}
\newcommand{\cmsq}{\rm cm^{-2}}
\newcommand{\intensity}{ergs cm$^{-2}$ s$^{-1}$ Hz$^{-1}$ sr$^{-1}$}

\title[J1211]{Cold gas and a Milky Way-type 2175 \AA$ $ bump in a metal-rich and highly depleted absorption system}
\author[Jingzhe Ma, Paul Caucal, Pasquier Noterdaeme, Jian Ge, J. Xavier Prochaska, Tuo Ji, Shaohua Zhang, Hadi Rahmani, Peng Jiang, Donald P. Schneider, Britt Lundgren, Isabelle P\^aris ]{Jingzhe Ma$^{1}$\thanks{E-mail:jingzhema@ufl.edu (JM)},  Paul Caucal$^{2,}$$^{11}$, Pasquier Noterdaeme$^{2}$, Jian Ge$^{1}$,  J. Xavier Prochaska$^{3}$, 
\newauthor Tuo Ji$^{4}$, Shaohua Zhang$^{4}$, Hadi Rahmani$^{5,}$$^{12}$, Peng Jiang$^{6}$, Donald P. Schneider$^{7,}$$^{8}$, 
\newauthor Britt Lundgren$^{9}$ and Isabelle P\^aris$^{10}$\\
$^{1}$Department of Astronomy, University of Florida, 211 Bryant Space Science Center, Gainesville, 32611, USA\\
$^{2}$Institut d'Astrophysique de Paris, CNRS-UPMC, UMR7095, 98bis bd Arago, 75014 Paris, France\\
$^{3}$Department of Astronomy and Astrophysics, UCO/Lick Observatory, University of California, 1156 High Street, Santa Cruz, 95064, USA\\
$^{4}$Polar Research Institute of China, 451 Jinqiao Road, Pudong, Shanghai 200136, China\\
$^{5}$Aix Marseille Universit\'e, CNRS, LAM (Laboratoire d'Astrophysique de Marseille) UMR 7326, 13388, Marseille, France\\
$^{6}$Key Laboratory for Research in Galaxies and Cosmology, University of Science and Technology of China, Chinese Academy of Sciences, Hefei, \\
          Anhui, 230026, China\\
$^{7}$Department of Astronomy and Astrophysics, The Pennsylvania State University, University Park, PA 16802, USA\\
$^{8}$Institute for Gravitation and the Cosmos, The Pennsylvania State University, University Park, PA 16802, USA\\
$^{9}$Department of Astronomy, University of Wisconsin - Madison, Madison, WI 53706, USA\\
$^{10}$Departamento de Astronom\'ia, Universidad de Chile, Casilla 36-D, Santiago, Chile\\
$^{11}$Master ICFP, D\'epartement de Physique, Ecole Normale Sup\'erieure, 24 rue Lhomond, 75231 Paris, France\\
$^{12}$School of Astronomy, Institute for Research in Fundamental Sciences (IPM), P.O. Box 19395-5531, Tehran, Iran}

\begin{document}

\pagerange{\pageref{firstpage}--\pageref{lastpage}} \pubyear{2002}

\maketitle

\label{firstpage}

\begin{abstract}
We report the detection of a strong Milky Way-type 2175 \AA$ $ extinction bump at $z$ = 2.1166 in the quasar spectrum towards SDSS J121143.42+083349.7 from the Sloan Digital Sky Survey (SDSS) Data Release 10. We conduct follow up observations with the Echelle Spectrograph and Imager (ESI) onboard the Keck-II telescope and the Ultraviolet and Visual Echelle Spectrograph (UVES) on the VLT. This 2175 \AA$ $ absorber is remarkable in that we simultaneously detect neutral carbon (C~{\sc i}), neutral chlorine (Cl~{\sc i}), and carbon monoxide (CO).  It also qualifies as a damped Lyman alpha system. The J1211+0833 absorber is found to be metal-rich and has a dust depletion pattern resembling that of the Milky Way disk clouds. We use the column densities of the C~{\sc i} fine structure states and the C~{\sc ii}/C~{\sc i} ratio (under the assumption of ionization equilibrium) to derive the temperature and volume density in the absorbing gas. A Cloudy photoionization model is constructed, which utilizes additional atoms/ions to constrain the physical conditions. The inferred physical conditions are consistent with a canonical cold (T $\sim$ 100 K) neutral medium with a high density ($n$(H~{\sc i}) $\sim$ 100 cm$^{-3}$) and a slightly higher pressure than the local interstellar medium. Given the simultaneous presence of C~{\sc i}, CO, and the 2175 \AA$ $ bump, combined with the high metallicity, high dust depletion level and overall low ionization state of the gas, the absorber towards J1211+0833 supports the scenario that the presence of the bump requires an evolved stellar population. 
\end{abstract}


\section{Introduction}

The most prominent feature in the interstellar extinction curve of the Milky Way (MW) is the broad absorption bump centered at rest-frame 2175 \AA$ $. The Large Magellanic Cloud (LMC) exhibits this feature as well, but the canonical Small Magellanic Cloud (SMC) extinction curve does not show a bump. The trend of decreasing bump strengths in the MW, LMC, and SMC corresponds to progressively lower metal abundances \citep{Fitzpatrick04} or different radiative environments (\citealt{Gordon97,Mattsson08}).   

Measuring the extinction curves in the Local Group by comparing stellar spectra is not feasible at larger distances. Instead, more luminous background sources are required to probe the gas and dust content beyond the local Universe.  Quasars, one of the brightest populations in the Universe, have been used as background sources to study absorption line systems. The intervening quasar absorption lines serve as a powerful tool for studying physical conditions, chemical abundances and gas kinematics in intervening absorbers, such as damped Lyman-$\alpha$ absorbers (DLAs), Mg~{\sc ii} absorbers, Ca~{\sc ii} and C~{\sc iv} absorbers. To date, a few dozens of the 2175 \AA$ $ bump detections have been reported in the spectra of intervening Mg~{\sc ii} absorbers (e.g., \citealt{Wang04,Srianand08,Jiang11,Zhou10,Zhang15}) and DLAs (e.g., \citealt{Wucknitz03,Junkkarinen04,Wang12}) towards quasars. Most of the detections are reported in individual absorption line systems. \cite{Jiang11} systematically selected thirty-nine 2175 \AA$ $ absorbers at $z$ $\sim$ 1-1.8 associated with strong Mg~{\sc ii} absorption lines from the Sloan Digital Sky Survey (SDSS; \citealt{York00}) Data Release 3, which is the largest sample to date. Most bumps in this sample resemble the relatively weak 2175 \AA$ $ bump in LMC supershell rather than the strong bump observed in the Milky Way. The 2175 \AA$ $ bump feature is also found in rare dust-rich and metal-strong DLAs (\citealt{Prochaska09,Wang12}), although DLAs are generally metal-poor systems with only a few percentage of the solar metal abundance (\citealt{Prochaska03,Kaplan10}). Recent years have seen the rapid growth of studies of Gamma-ray burst (GRB) afterglows which are promising in probing even higher redshift absorbing systems.  A few cases of GRB afterglow spectra also reveal the 2175 \AA$ $ bump from intervening systems or from gas in the GRB host galaxies (e.g., \citealt{Prochaska09,Eliasdottir09,Zafar12}). 

The nature of the 2175 \AA$ $ bump has been investigated for almost  50 years since its first discovery \citep{Stecher65}. \cite{Noll07} found a correlation between heavy reddening and the presence of the bump. Several studies (e.g., \citealt{Noterdaeme09,Jiang10b}) found that a high dust depletion level (i.e., [Fe/Zn] $\sim$ -1.5) is required to produce a prominent 2175 \AA$ $ bump. Simultaneous detections of metal lines, neutral carbon (\citealt{Eliasdottir09,Noterdaeme09}), and molecular gas (\citealt{Prochaska09,Noterdaeme09}) with the 2175 \AA$ $ bump tend to suggest a correlation between those ingredients and the presence of the bump. The exact origin or carrier of the 2175 \AA$ $ bump remains an open question, although candidates have been suggested to be some form of graphitic carbon, most likely polycyclic aromatic hydrocarbons (PAHs; \citealt{Draine03}).

We initiated a project to search for 2175 \AA$ $ bumps in SDSS quasar spectra, yielding a large sample of 2175 \AA$ $ absorbers (Zhang et al. in prep).  From this survey, an intervening system towards SDSS J121143.42+083349.7 (hereafter J1211+0833\footnote{J1211+0833 is used to name the quasar. We refer to the absorber as ``the (intervening) absorber/system towards J1211+0833" or ``the J1211+0833 absorber" throughout the paper.}) stands out as a remarkable 2175 \AA$ $ absorber with a significant absorption feature and rich metal lines.  Along with the bump in the same absorber, we have simultaneously detected C~{\sc i} fine structure lines,  CO absorption bands, and molecular hydrogen hand in hand with neutral chlorine. 

C~{\sc i} fine structure transitions have been recognized and used  to probe cold gas in high-redshift universe. The ground state of C~{\sc i} is split into three fine structure levels referred to as C~{\sc i}, C~{\sc i}$^*$, and C~{\sc i}$^{**}$. They are sensitive probes due to the fact that the energy separations between excited states and the ground state are only 23.6 K and 62.4 K, respectively.  Physical conditions of the absorber (i.e., temperature, density, and pressure) can be implied by the relative excitation of the C~{\sc i} fine structure states \citep{Jenkins79}. In the Milky Way, surveys of interstellar C~{\sc i} fine structure excitations (\citealt{Jenkins01,Jenkins07, Jenkins11}) find that in the local interstellar medium (ISM) the gas pressure is mostly between 3 $<$ log($P$/k) $<$4 cm$^{-3}$ K.  C~{\sc i} analysis has revealed the physical conditions in several high-z DLAs (e.g., \citealt{Ge97a,Srianand05,Noterdaeme07}) together with detections of H$_2$. CO is detected in a few intervening systems (e.g., \citealt{Srianand08,Prochaska09,Noterdaeme09,Jorgenson10} ). Analyzing the excitation/de-excitation of the molecular and atomic species together provides detailed physical conditions of the absorbing gas and much needed insight into the properties of the intervening galaxy (i.e., its chemical abundances, dust content, and star formation rate). 

This paper examines the properties of this particular 2175 \AA$ $ absorber, and is organized as follows. In Section \ref{sec:observation}, we describe target selection, observations, and data reduction conducted on J1211+0833.  In Section \ref{sec:bump}, we derive the bump strength, bump width and the extinction curve.  In Section \ref{sec:N}, we show the velocity profiles of absorption lines and column density measurements. Gas-phase metal abundances and dust depletion pattern are derived in Section \ref{sec:depletion}. In the next section, we report the detections of C~{\sc i}, Cl~{\sc i},  and CO. We derive the physical conditions through the C~{\sc i} method and also construct a Cloudy photoionization model. The results are discussed and summarized in Section \ref{sec:summary}.

\section[]{Target Selection, Observations, and Data Reduction}
\label{sec:observation}

J1211+0833 was originally selected from the tenth data release (DR10) of the SDSS-III's Baryon Oscillation Spectroscopic Survey (BOSS; \citealt{Dawson13,Eisenstein11,Ross12}). It was targeted for spectroscopic observations on 2012 January 19. The BOSS spectra \citep{Smee13} cover a wavelength range of  3650-10400 \AA$ $ with a spectral resolution of R $\sim$ 2000. The quasar redshift is $z_{qso}$=2.483 \citep{Bolton12,Paris14}.  A significantly broad absorption feature is clearly visible in the spectrum (Fig. \ref{fig:bumpfitting}), which is later confirmed as the 2175 \AA$ $ extinction bump. The point-spread-function $ugriz$ magnitudes for J1211+0833 are $u$ = 21.10 $\pm$ 0.10, $g$ = 19.63 $\pm$ 0.01, $r$ = 19.47 $\pm$ 0.02, $i$ = 19.15 $\pm$ 0.02, and $z$ = 18.41 $\pm$ 0.03. 

We conducted follow-up observations of J1211+0833 with the Echelle Spectrographs and Imager (ESI; \citealt{Sheinis02}) onboard the Keck II telescope in order to reveal at higher spectral resolution the detailed properties of the ISM towards the quasar J1211+0833.  The spectra were obtained on 2013 March 8 with two exposures of 1800 s and 1500 s, using the 0.75$\arcsec$ slit with a corresponding resolution of R $\sim$ 5400 . The seeing conditions during observation varied from 1.5$\arcsec$ to 2$\arcsec$. The data were reduced, calibrated, and combined by the ESI Echelle Data Reduction package \footnote{http://www2.keck.hawaii.edu/inst/esi/ESIRedux/}. The final combined spectrum covers a wavelength range of 3900 \AA$ $ to 11715 \AA$ $ with a median S/N $\sim$ 9 per pixel around 6500 \AA$ $. The standard star G19B2B was used for relative flux calibration. 

J1211+0833 was also observed with the Ultraviolet and Visible Echelle Spectrograph (UVES; Dekker et al. 2000) mounted on the VLT on March 25 and April 24, 2014 (Program ID: 093.A-0126) under good seeing conditions (FWHM $<$ 0.7 $\arcsec$). Four 4350 s exposures were taken simultaneously with the blue and red arms. We used a 0.9 \arcsec slit and a pixel binning of 2 $\times$ 2, leading to a resolving power of R $\sim$ 54000 and a mean S/N of $\sim$ 6. The data were reduced using the UVES pipeline (Ballester et al. 2000). The individual exposures were scaled and combined together, weighted by the inverse variance in each pixel. The cosmic ray impact residuals were rejected at the same time. The actual wavelengths covered in the reduced spectra are from 3288 \AA$ $ to 6648 \AA$ $ with two gaps between 4525 - 4621 \AA$ $ and 5600 - 5647 \AA$ $.

\section{The 2175 \AA$ $ extinction bump}
\label{sec:bump}

To extract the properties of the 2175 \AA$ $ extinction bump towards J1211+0833, we adopt the \cite{Fitzpatrick90} parametrization (FM parameterization) of the optical/UV extinction curve. The  extinction curve in the rest frame of the absorber can be described as a combination of a linear component, characterizing the underling extinction, and a Drude profile, modeling the potential 2175 \AA$ $ bump.  The extinction curve is parameterized as 
\begin{equation}
 A(\lambda)=c_1+c_2x+c_3D(x,x_0,\gamma),   
\end{equation}
where $x$ = $\lambda^{-1}$, and the Drude profile $D(x,x_0,\gamma)$ has the form of 
\begin{equation}
D(x,x_0,\gamma)=\frac{x^2}{(x^2-x_0^2)^2+x^2\gamma^2},
\end{equation}
where $x_0$ and $\gamma$ are the bump peak position and bump width (FWHM) of the Drude profile, respectively. The UV linear component is set by the slope $c_2$ and intercept $c_1$. The derived extinction curve is not the absolute extinction curve normalized by $E(B-V)$ but in a relative sense. The strength of the bump can be defined using these parameters: $A_{bump}=\pi c_3/(2\gamma)$ measures the area of the bump \footnote{The A$_{bump}$ defined in this paper is slightly different from that in \cite{Fitzpatrick07} since their parameterized extinction curves are normalized by $E(B-V)$. Therefore A$_{bump}$ = $E(B-V)$ $\times$ A$^*_{bump}$ where A$^*_{bump}$ is the bump strength defined in \cite{Fitzpatrick07}.}.  We adopt the SDSS DR7 composite quasar spectrum \citep{Jiang11} as the intrinsic spectrum for J1211+0833, which is an update of the SDSS quasar composite spectrum of \cite{VandenBerk01} by median combining 105783 quasar spectra in the SDSS DR7 \citep{Schneider10}. The composite spectrum is reddened using the FM parameterized extinction curve at $z_{abs}$=2.1166 to form the model spectrum. The absorber's redshift is based upon the strong absorption lines such as Mg~{\sc ii} in the Keck spectrum. To optimize the continuum fitting, strong emission lines such as Ly$\alpha$, Si~{\sc iv}, C~{\sc iv}, C~{\sc iii}, and Mg~{\sc ii} and known absorption lines are masked out. 

\begin{figure*}
\centering
{\includegraphics[width=15cm, height=9cm]{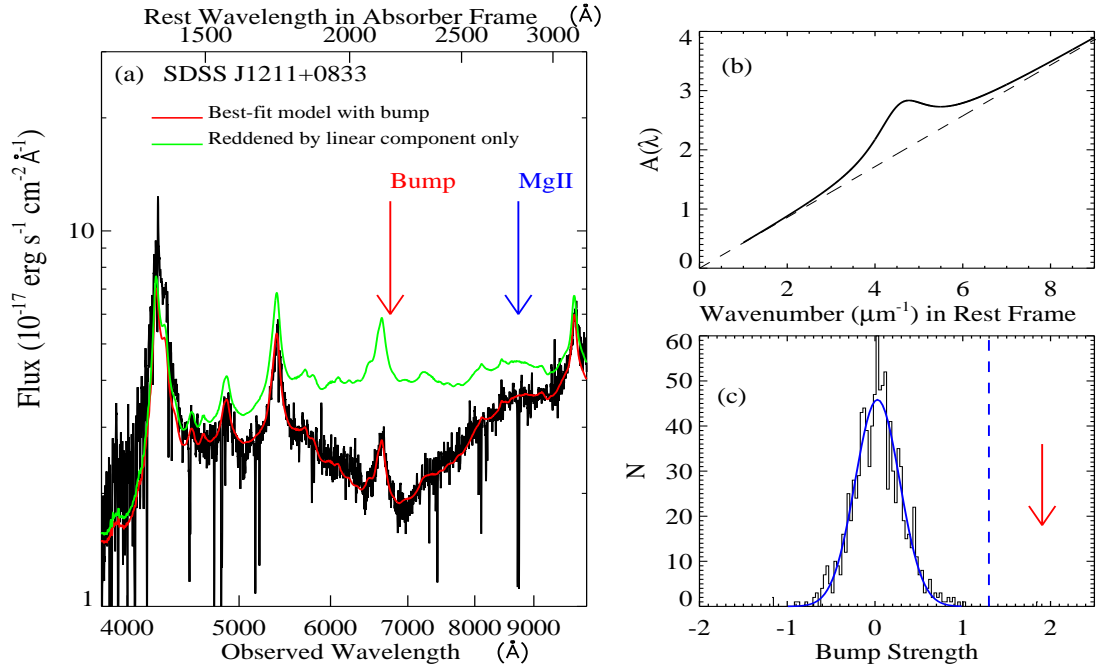}}   
\caption{The 2175 \AA$ $ bump fitting results for the absorber towards J1211+0833.  (a) The black spectrum is the observed quasar spectrum retrieved from BOSS DR10. The red curve is the best-fit model with the extinction bump whose central position is indicated by the red arrow and the blue arrow points to the Mg~{\sc ii} absorption lines. The green curve is the best-fit model reddened by the linear component only. (b) The best-fit extinction curve with the constraint that the extinction should be zero at $\lambda \rightarrow \infty$. (c) The bump strength distribution of the control sample fit by a Gaussian curve in blue. The vertical dashed blue line indicates 5$\sigma$ and the red arrow denotes the bump strength towards J1211+0833. }
\label{fig:bumpfitting}
\end{figure*}

\begin{figure*}
\centering
{\includegraphics[width=16cm, height=9cm]{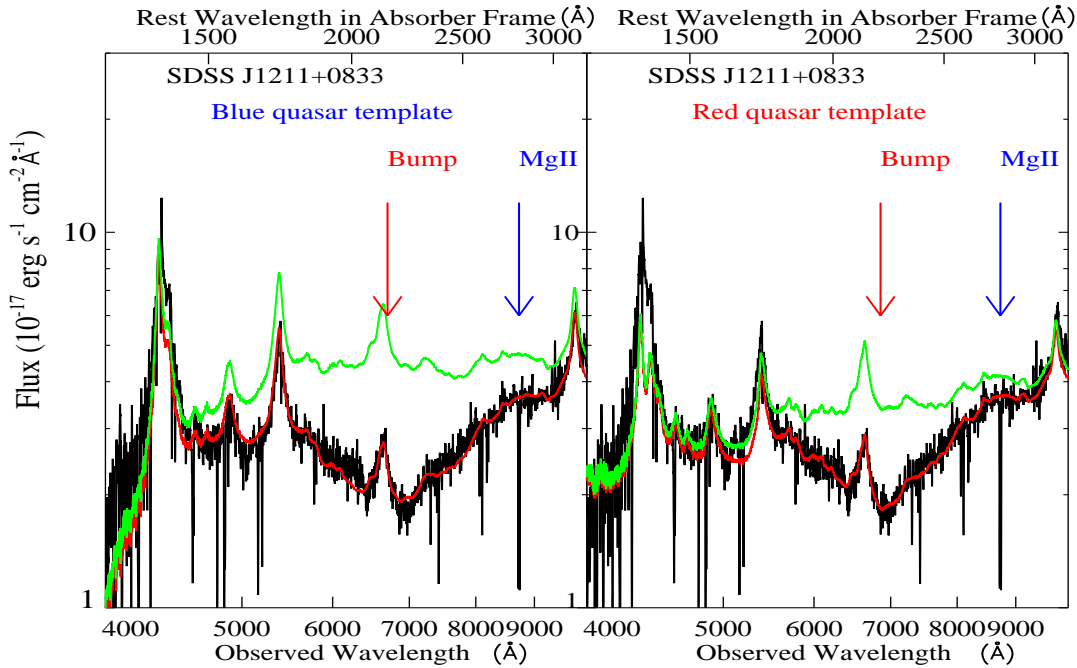}}   
\caption{The 2175 \AA$ $ bump fitting results using the blue (left) and red (right) quasar templates. The black spectrum is the observed quasar spectrum retrieved from BOSS DR10. The red curve is the best-fit model with the extinction bump the central position of which is indicated by the red arrow and the blue arrow points to the Mg~{\sc ii} absorption lines. The green curve is the best-fit model reddened by the linear component only. The 2175 \AA$ $ bump is present even in the extreme case of the red template.}
\label{fig:templatefitting}
\end{figure*}

We compare the model spectrum with the observed BOSS spectrum which has been corrected for Galactic extinction by using the dust map of \cite{Schlegel98} to find the best-fit.  Fig. \ref{fig:bumpfitting} (a) demonstrates the fitting result with the black spectrum being the observed spectrum and the red curve representing the best-fit model. The best-fit parameters are $c_1$ = -0.13, $c_2$ = 0.43, $c_3$ = 1.78, $x_0$ = 4.6, $\gamma$ = 1.47, and the bump strength $A_{bump}$ = 1.90 with $\chi^2$ = 1.21. In contrast, the green curve shows the composite quasar spectrum reddened by the best-fit linear component without an extinction bump, which explicitly illustrates the existence of a strong 2175 \AA$ $ bump.  As the intrinsic spectrum of J1211+0833 is unknown, we should also consider the possibility of an intrinsically red or blue quasar spectrum that potentially provides a better fit. We select the bluest and reddest 20\% quasars respectively from the SDSS DR7 quasar sample according to spectral index. The blue and red quasar templates are constructed by median combining those selected quasar spectra. We perform the same fitting method using the blue and red templates (Fig.\ref{fig:templatefitting}) as the intrinsic quasar spectra for J1211+0833, respectively.  Comparing with the composite spectrum, an intrinsically blue or red quasar template provides an inferior fit in terms of representing the observed spectrum, which justifies the usage of the composite spectrum as a better model representative for J1211+0833. The bump strength and width increase to $A_{bump}$ = 2.42 and $\gamma$ = 1.67 in the case of the blue template and drop to  $A_{bump}$ = 1.15 and $\gamma$ = 1.11 for the red template. 

The strength of the 2175 \AA$ $ bump can appear to be stronger or weaker due to the intrinsic variation of quasar spectra, which in some cases  imitates an extinction bump \citep{Pitman00}. We therefore utilize the simulation technique developed by \cite{Jiang10a} to set the criteria for the significance of the bump.  The first step is to select a control sample of SDSS quasar spectra (sample size $\sim$ 1000) at a similar redshift to the quasar of interest. For J1211+0833, we select quasars with redshifts in the range of $z_{qso}$-0.05$<$ $z$ $<$ $z_{qso}$+0.05 and $i$-band S/N $\geq$ 5 as its control sample. Then we fit each of them by reddening the composite quasar spectrum with a parameterized extinction curve at the redshift of the bump absorber of interest.  The parameters $x_0$ and $\gamma$ in the Drude profile are held fixed to the best values derived from the fitting of the quasar of interest.  The expected bump strength distribution is a Gaussian assuming random fluctuations of continuum of each spectrum in the control sample.  Bumps that significantly deviate from the Gaussian distribution are defined to have statistical significance. The histogram in Fig. \ref{fig:bumpfitting} (c) shows that the bump towards J1211+0833 lies at a statistical confidence level greater than 5$\sigma$ indicated by the red arrow. 

\begin{figure*}
\centering
{\includegraphics[width=13cm, height=9cm]{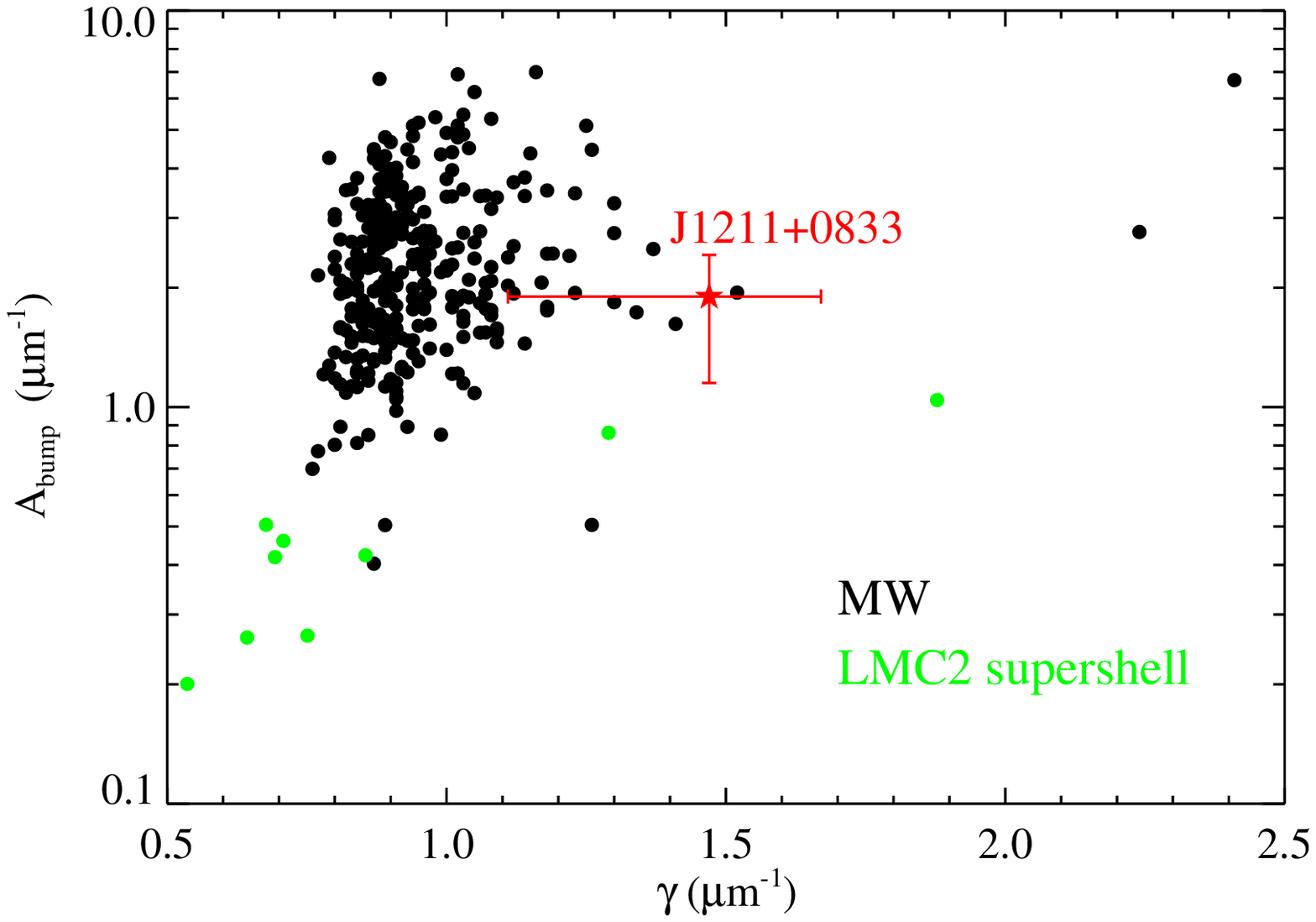}}  
\caption{The comparison of 2175 {\AA} bump strengths $A_{bump}$ and bump widths $\gamma$ with extinction bumps in the MW \citep{Fitzpatrick07} and LMC2 \citep{Gordon03} supershell denoted with black and green circles, respectively. The absorber towards J1211+0833 is denoted as the red star. The error bars reflect the uncertainty induced by the blue and red template fitting.}
\label{fig:MW_LMC}
\end{figure*}

The best studied 2175 \AA$ $ bumps are those along lines of sight in our Milky Way and towards the LMC. We compare the bump strengths in the $A_{bump}-\gamma$ space (Fig. \ref{fig:MW_LMC}) where the black filled circles are the bumps in the MW \citep{Fitzpatrick07} and the green circles are LMC2 supershell bumps \citep{Gordon03}. The J1211+0833 absorber possesses a typical MW-type bump strength but is wider than the majority.  Does this indicate a mature environment resembling the MW has enabled the formation of such a strong extinction bump at the absorber's redshift of $z$ = 2.12? We resort to the column density measurements, metal abundances, and dust depletion pattern analysis to explore the physical and chemical environment in the absorber. 

\section{Absorption Lines and Column Density Measurements}
\label{sec:N}

\subsection{Metal lines in Keck/ESI}

\begin{figure*}
\centering
{\includegraphics[width=16cm, height=23cm]{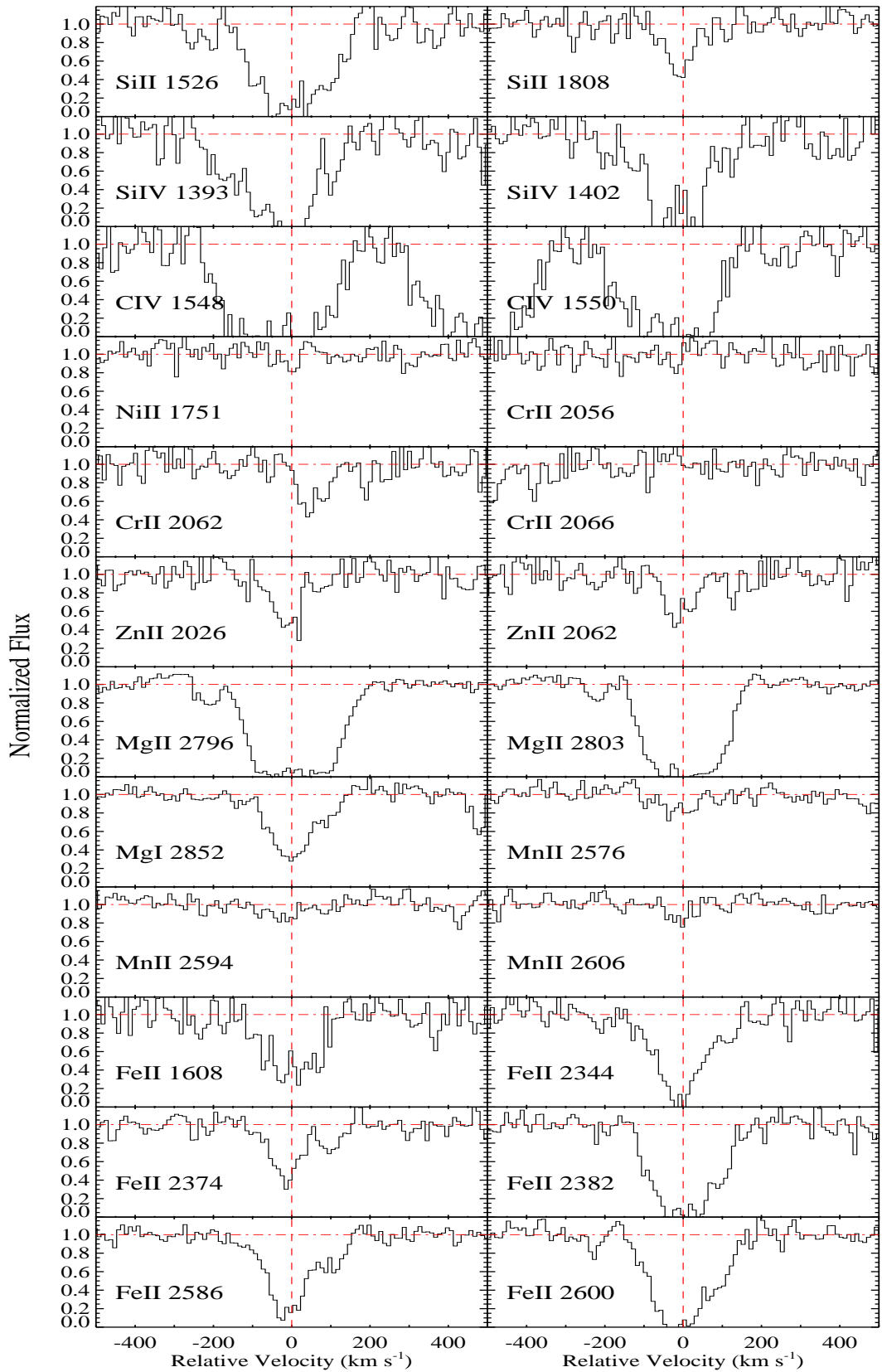}}   
\caption{Metal absorption lines detected in the Keck/ESI for the absorber at $z$=2.1166 towards J1211+0833.  }
\label{fig:ESIlines}
\end{figure*}

Numerous narrow absorption lines are detected in the Keck/ESI spectra of J1211+0833, enabling measurements of gas-phase column densities of many ions. The detected metal lines are listed in Table \ref{tab:ESIlines} and the absorption lines of interest are plotted in velocity space (i.e., normalized flux vs. relative velocity) in Fig. \ref{fig:ESIlines}. 

\begin{figure*}
\centering
{\includegraphics[width=12.5cm, height=9cm]{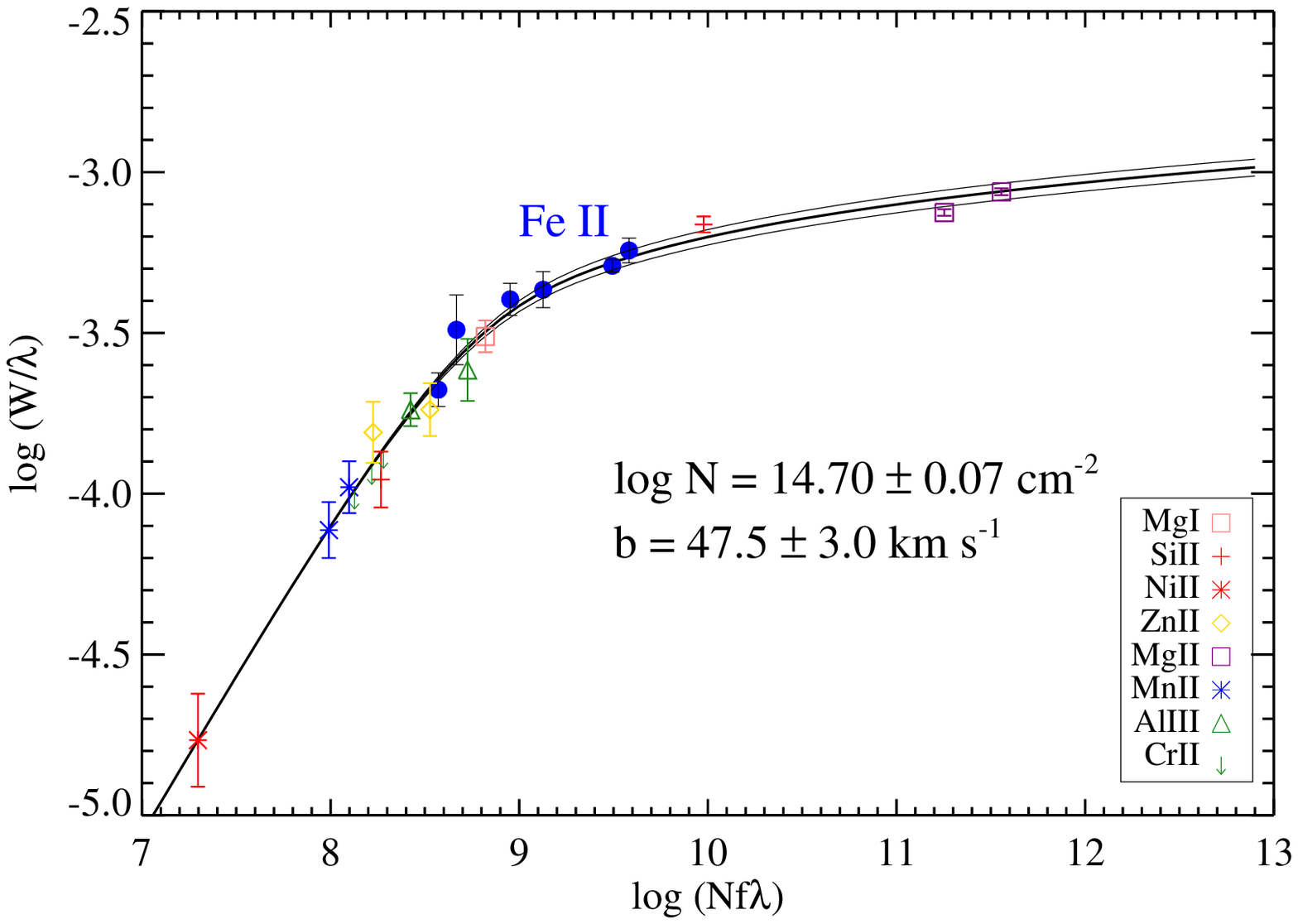}} 
\caption{The COG using six Fe~{\sc ii} absorption lines from Keck/ESI. The three curves represent the best-fit and corresponding 1$\sigma$ confidence level. The column densities of other atoms/ions are derived based on the best-fit COG and the measured EWs. }
\label{fig:cog}
\end{figure*}

To measure column densities of metal absorption lines in the 2175 \AA$ $ absorption system, we normalize the observed Keck spectrum locally by the quasar continuum, which is fitted  by polynomials with the absorption feature masked during the fitting. In the framework of column density measurements, the Apparent Optical Depth Method (AODM; \citealt{Savage91}) is often applied to resolved spectra and the Curve of Growth (COG) method \citep{Jenkins86}  is commonly used in unresolved cases. As the Keck/ESI spectrum for J1211+0833 is partially resolved, we adopt both methods to derive the column densities as a consistency check.  The AODM is based upon the connection between optical depth and column density: the column densities are measured by integrating the optical depth over the velocity range of the spectra covering the detected transitions.  The column densities derived from this method are lower limits for the saturated lines.  For the COG method, the Fe~{\sc ii} $\lambda\lambda$ 1608, 2344, 2374, 2382, 2586 and 2600 absorption lines, arising from the ground level $Fe^+$, are used to construct the COG in Fig. \ref{fig:cog}. They all have the same column density $N$(Fe~{\sc ii}) and the Doppler parameter $b$. The rest frame equivalent widths (EWs) are measured by integrating over a typical velocity interval of [-400, 400]  ${\rm km}$ $ {\rm s}^{-1}$ to include the absorption feature as well as to avoid contamination from other absorption. The best-fit COG is obtained by finding the best match between theoretical COG calculated with various $b$ values and the measured EW of the six Fe~{\sc ii} absorption lines. We explore the $b -N$(Fe~{\sc ii}) parameter space based on $\chi^2$ minimization to find the best fit parameters  log $N$(Fe~{\sc ii}) = 14.70 $\pm$ 0.07 ${\rm cm^{-2}}$ and $b$ = 47.5 $\pm$ 3.0 ${\rm km}$ $ {\rm s^{-1}}$. The uncertainties quoted refer to the 1$\sigma$ confidence level. The column densities of other atoms/ions are derived based on the best-fit COG and the measured EWs. For species with multiple absorption lines, the column densities are derived by simultaneously fitting all the available transitions.  The $N_{AODM}$ and $N_{COG}$ values are in good agreement for the unsaturated lines.

%%%%%%%%%%%%
%ESI metal lines
%%%%%%%%%%%%
\begin{table*}
\caption{Column density measurements of the system towards J1211+0833 using the Keck/ESI data. The central wavelengths of the transitions and the equivalent widths are reported in the absorber rest frame. We adopt the oscillator strengths, $f$, from the Atomic Data compiled by \citet{Morton03}. Saturated lines have lower limits placed on their column densities derived from the AODM. Undetected lines have 3$\sigma$ upper limits on equivalent widths and hence column densities from the COG method. }
\begin{tabular}{lccccc}
\hline
\hline
{\parbox[t]{15mm}{\centering Ion}} &                          
{\parbox[t]{25mm}{\centering Transition}} &                            
{\parbox[t]{15mm}{\centering $f$}} &  
{\parbox[t]{15mm}{\centering EW }} &
{\parbox[t]{20mm}{\centering log $N_{AODM}$}} &
{\parbox[t]{20mm}{\centering log $N_{COG}$}}\\          
     &(\AA)  &   &  (\AA)  &  (cm$^{-2}$)   & (cm$^{-2}$)  \\
\hline
FeII  1608  &  1608.4511   &  0.0577   &  0.52 $\pm$ 0.13   & 14.74 $\pm$ 0.07  & 14.70 $\pm$ 0.07   \\
FeII  2344  &  2344.2139   &  0.1140   &  1.01 $\pm$ 0.13   & 14.52 $\pm$ 0.03  &  ...   \\
FeII  2374  &  2374.4612   &  0.0313   &  0.50 $\pm$ 0.06   & 14.61 $\pm$ 0.06  & ... \\
FeII  2382  &  2382.7652   &  0.3200   &  1.36 $\pm$ 0.12   & 14.59 $\pm$ 0.01  & ...  \\
FeII  2586  &  2586.6500   &  0.0691   &  1.04 $\pm$ 0.06   & 14.62 $\pm$ 0.03  & ...  \\ 
FeII  2600  &  2600.1729   &  0.2390   &  1.33 $\pm$ 0.10   & 14.73 $\pm$ 0.01  & ...  \\
MgI   2852  &  2852.9631   &  1.8300   &  0.88 $\pm$ 0.10   & 12.98 $\pm$ 0.03  & 13.08 $\pm$ 0.10   \\
MgII  2796  &  2796.3553   &  0.6155   &  2.43 $\pm$ 0.06   & $>$14.20          & 15.97 $\pm$ 0.09   \\
MgII  2803  &  2803.5324   &  0.3058   &  2.10 $\pm$ 0.05   & $>$14.47          & ... \\
SiII  1526  &  1526.7070   &  0.1330   &  1.05 $\pm$ 0.06   & $>$14.95   & 15.65 $\pm$ 0.07   \\
SiII  1808  &  1808.0129   &  0.0021   &  0.20 $\pm$ 0.04   & 15.64 $\pm$ 0.10  & ...   \\
NiII  1751  &  1751.9157   &  0.0277   &  0.03 $\pm$ 0.01   & 13.68 $\pm$ 0.69  & 13.61 $\pm$ 0.18   \\
ZnII  2026  &  2026.1370   &  0.5010   &  0.37 $\pm$ 0.07   & 13.43 $\pm$ 0.08  & 13.51 $\pm$0.09   \\
ZnII  2062  &  2062.6604   &  0.2460   &  0.32 $\pm$ 0.07   & 13.61 $\pm$ 0.11  & ...   \\
MnII  2576  &  2576.8770   &  0.3610   &  0.27 $\pm$ 0.05   & 13.14 $\pm$ 0.11  & 13.09 $\pm$ 0.06   \\
MnII  2594  &  2594.4990   &  0.2800   &  0.20 $\pm$ 0.04   & 13.10 $\pm$ 0.14  & ... \\
MnII  2606  &  2606.4620   &  0.1980   &  0.12 $\pm$ 0.04   & 12.69 $\pm$ 0.48  & ...  \\
CrII  2056  &  2056.2569   &  0.1030   &  $<$0.25           &                 & $<$13.95           \\
CrII  2062  &  2062.2361   &  0.0759   &  $<$0.43           &                 & ...       \\
CrII  2066  &  2066.1640   &  0.0512   &  $<$0.28           &                 & ...         \\           
AlIII 1854  &  1854.7184   &  0.5590   &  0.45 $\pm$ 0.10   & 13.76 $\pm$ 0.04  & 13.72 $\pm$ 0.07   \\ 
AlIII 1862  &  1862.7910   &  0.2780   &  0.34 $\pm$ 0.04   & 13.79 $\pm$ 0.08  & ...  \\
SiIV 1393  &  1393.7602     &  0.5130 &                               & $>$14.49                &                               \\
SiIV 1402  &  1402.7729    &  0.2540  &                               & $>$14.71                 &                              \\
CIV  1548  &  1548.1950   &  0.1908   &                               & $>$15.02                 &                               \\
CIV  1550  &  1550.7700  &  0.0952   &                                & $>$15.27               &                                \\
\hline\\[0.02mm]
\end{tabular}
\label{tab:ESIlines}
\end{table*}

\subsection{Metal lines in VLT/UVES and velocity structure}

\begin{figure}
{\includegraphics[width=8cm, height=10cm]{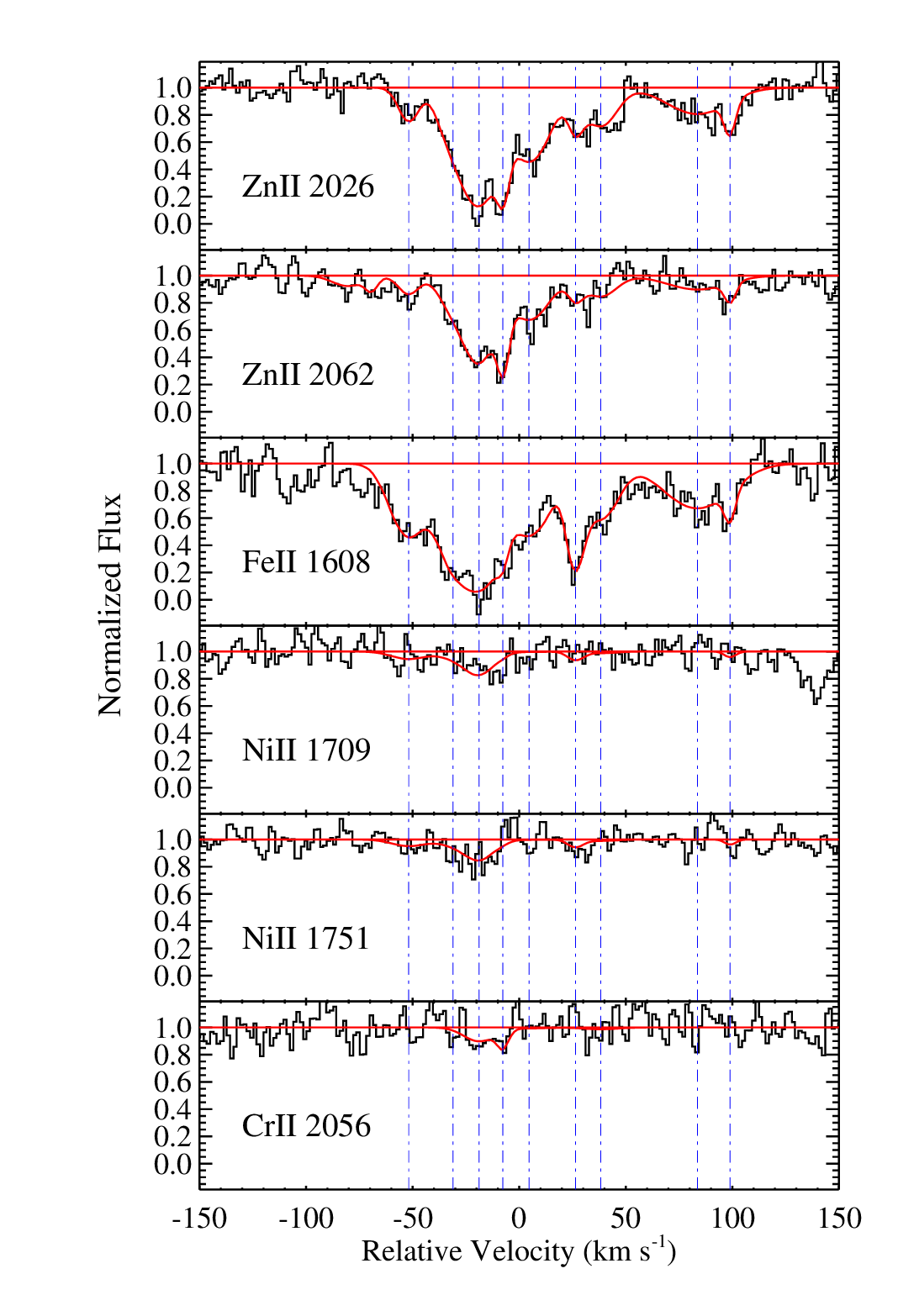}}   
\caption{Simultaneous multi-component Voigt-profile fitting with VPFIT on the VLT/UVES data. The original spectra are in black and the fitted profiles are in red. A minimum of 9 velocity components indicated by the vertical lines are required to optimally fit the absorption features. }
\label{fig:UVESlines}
\end{figure}

The Doppler parameter derived from metal lines in the Keck/ESI spectrum implies there is blending, which is confirmed by the high resolution (5.5 \kms) VLT/UVES spectrum.  The metal absorption lines are modeled by Voigt profiles using the VPFIT package (version 9.5) \footnote{http://www.ast.cam.ac.uk/$\sim$rfc/vpfit.html}.  VPFIT utilizes a $\chi^2$-minimization algorithm for multi-component Voigt-profile fitting to derive absorption redshifts, Doppler parameters, and column densities taking into account the instrumental resolution. A set of different singly ionized lines covered in VLT/UVES (i.e., Zn~{\sc ii}, Fe~{\sc ii}, Ni~{\sc ii}, Cr~{\sc ii}) are fitted simultaneously assuming they are kinematically associated with the same gas cloud.  The redshift and $b$ parameter for each component are tied to each other. We find that a minimum of nine velocity components are required to optimally fit the absorption features in strong Zn~{\sc ii} and Fe~{\sc ii} lines (Fig. \ref{fig:UVESlines}). We number the components from blue to red as components 1 to 9 at the following velocities relative to $z_{abs}$ = 2.1166 for $v$ = 0 \kms (the center identified in Keck/ESI): $v$ $\sim$ -52 \kms, -31 \kms, -19 \kms, -8 \kms, +5 \kms, +26 \kms, +38 \kms, +84 \kms, and +99 \kms. Ni~{\sc ii} and Cr~{\sc ii} line profiles are attributed to the primary components 3 and 4 at $z$ = 2.116404 and $z$ = 2.116520, respectively. The nine components span a velocity interval of $\Delta$$v$ $\sim$ 150 \kms. Note that the correspondence between $\Delta$$v$ and $b$ (from the curve of growth on ESI) matches very well the relation from \cite{Noterdaeme14}: $\Delta$$v$ = 2.21$b_{eff}$ + 0.02$b_{eff}^2$. The profiles exhibit an edge-leading asymmetry with the strongest feature in the blue end. The asymmetric shape is consistent with the prediction by a simple model of a rotating disk \citep{Prochaska97}.

The column densities in each component and associated errors are provided by VPFIT and are reported in Table \ref{tab:UVESlines}. The total column densities for each ion summed over all the components, log $N$(Zn~{\sc ii}) = 13.58 $\pm$ 0.06, log $N$(Fe~{\sc ii}) = 14.77 $\pm$ 0.03, log $N$(Ni~{\sc ii}) = 13.62 $\pm$ 0.08, and log $N$(Cr~{\sc ii}) = 12.78 $\pm$ 0.14 $\cmsq$, are consistent with the measurements from Keck/ESI. Potential contamination to Zn~{\sc ii} column density from Cr~{\sc ii} $\lambda$2026 and Cr~{\sc ii} $\lambda$2062 lines is taken into account in the simultaneous fitting. Note that the $b$ values obtained from the fit can be smaller than the velocity resolution of the instrument because the Doppler parameters are constrained by simultaneously fitting several transitions with a range of oscillator strengths (i.e., probing different regimes along the curve of growth). The real resolution can be better than the nominal resolution we used in the fitting because the seeing is less than the slit width. This may well explain the small $b$-values we obtained.

%%%%%%%%%%%%
%UVES metal lines
%%%%%%%%%%%%

\begin{table}
\centering
\caption{Doppler parameters and column densities for each component measured using the VLT/UVES data.}
\begin{tabular}{cccccc}
\hline\hline
Comp. & $z_{abs}$ & Ion &  $b$ (km s$^{-1}$) &  log $N$ ($\mathrm{cm^{-2}}$)      \\
\hline
1   &2.116061  & Zn~{\sc ii}   &5.58       &12.09$\pm$0.07    \\
    &          & Fe~{\sc ii}   &10.60      &13.79$\pm$0.04    \\
    &          & Ni~{\sc ii}   &10.60      &12.88$\pm$0.15    \\
    &          & Cr~{\sc ii}  &                &                   \\
\hline
2   &2.116277  & Zn~{\sc ii}   &8.20$\pm$1.12   &12.49$\pm$0.09    \\
    &          & Fe~{\sc ii}   &8.87            &13.93$\pm$0.07    \\
    &          & Ni~{\sc ii}   &8.66            &12.67$\pm$0.28    \\
    &          & Cr~{\sc ii}   &                &                   \\
\hline
3   &2.116404  & Zn~{\sc ii}   &8.38$\pm$1.28   &13.07$\pm$0.05     \\
    &          & Fe~{\sc ii}   &9.06            &14.25$\pm$0.08     \\
    &          & Ni~{\sc ii}   &8.84            &13.32$\pm$0.08     \\
    &          & Cr~{\sc ii}   &9.39            &12.53$\pm$0.13     \\
\hline
4   &2.116520  & Zn~{\sc ii}   &1.80$\pm$0.68   &12.91$\pm$0.23    \\
    &          & Fe~{\sc ii}   &1.95            &13.36$\pm$0.30    \\
    &          & Ni~{\sc ii}   &1.90            &11.79$\pm$1.37    \\
    &          & Cr~{\sc ii}   &2.02            &12.27$\pm$0.17    \\
\hline
5   &2.116648  & Zn~{\sc ii}   &12.02$\pm$1.18  &12.81$\pm$0.03    \\
    &          & Fe~{\sc ii}   &13.01           &13.86$\pm$0.04    \\
    &          & Ni~{\sc ii}   &12.69           &11.70$\pm$2.58    \\
    &          & Cr~{\sc ii}   &13.48           &11.47$\pm$1.64    \\
\hline
6   &2.116875  & Zn~{\sc ii}   &3.74$\pm$0.71   &12.04$\pm$0.10    \\
    &          & Fe~{\sc ii}   &4.05            &13.74$\pm$0.06    \\
    &          & Ni~{\sc ii}   &3.95            &12.61$\pm$0.21    \\
    &          & Cr~{\sc ii}  &                &                   \\
\hline
7   &2.116998  & Zn~{\sc ii}   &9.53$\pm$1.33   &12.35$\pm$0.06    \\
    &          & Fe~{\sc ii}   &10.31           &13.61$\pm$0.06    \\
    &          & Ni~{\sc ii}   &10.06           &12.19$\pm$0.75    \\
    &          & Cr~{\sc ii}   &10.69           &11.68$\pm$0.99    \\
\hline
8   &2.117469  & Zn~{\sc ii}   &18.94$\pm$1.93  &12.46$\pm$0.06    \\
    &          & Fe~{\sc ii}   &20.49           &13.78$\pm$0.05    \\
    &          & Ni~{\sc ii}   &                &                   \\
    &          & Cr~{\sc ii}   &                &                   \\
\hline
9   &2.117628  & Zn~{\sc ii}   &1.62$\pm$1.53   &11.97$\pm$0.12    \\
    &          & Fe~{\sc ii}   &1.75            &12.98$\pm$0.14    \\
    &          & Ni~{\sc ii}   &1.71            &12.30$\pm$0.36    \\
    &          & Cr~{\sc ii}   &                &                   \\
\hline
total &        & Zn~{\sc ii}   &                &13.58$\pm$0.06 \\ 
      &        & Fe~{\sc ii}   &                &14.77$\pm$0.03  \\
      &        & Ni~{\sc ii}   &                &13.62$\pm$0.08  \\
      &        & Cr~{\sc ii}   &                &12.78$\pm$0.14  \\
\hline
\end{tabular}
\label{tab:UVESlines}
\end{table}

\subsection{Lyman-$\alpha$ profile fitting}

\begin{figure*}
\centering
{\includegraphics[width=13cm, height=8cm]{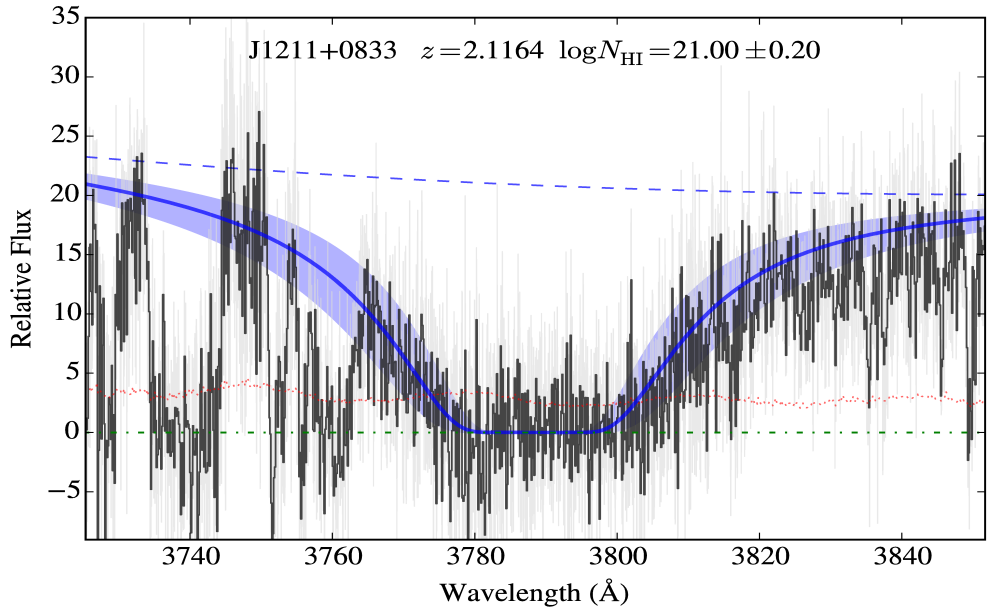}}
\caption{DLA profile fitting on the VLT/UVES spectrum (grey). The spectrum plotted in black is the original spectrum smoothed using 5-pixel boxcar and the red dotted line represents the associated flux errors. The blue line is the best-fit Voigt profile with a hydrogen column density of log $N$(H~{\sc i}) = 21.00 $\cmsq$ and the estimated 1 $\sigma$ uncertainty is denoted by the blue shade. The blue dashed line is the associated continuum. }
\label{fig:dlafitting}
\end{figure*}

The Ly$\alpha$ absorption line of the system towards J1211+0833, covered by the VLT/UVES spectrum, is fitted by a Voigt profile using $x\_fitdla$ from the XIDL package \footnote{http://www.ucolick.org/$\sim$xavier/IDL/index.html}. We adopt the best-fit Voigt profile (Fig. \ref{fig:dlafitting}) with a neutral hydrogen column density of log $N$(H~{\sc i}) = 21.00 $\pm$ 0.20 ${\rm cm^{-2}}$. The error reflects the systematic uncertainty induced by the continuum level plus small uncertainty in the fitting. We also utilize VPFIT to perform the Ly-$\alpha$ profile fitting which is based on normalized flux. We use both as a sanity check. The resulting best-fit log $N$(H~{\sc i}) is 0.05 dex higher and well within the estimated error.  The system towards J1211+0833 is therefore classified into a DLA \citep{Wolfe86}. With the hydrogen column density available, we are able to derive the absolute metal abundances.

%%%%%%%%%%%%%%%
%Lyman-alpha profile fitting
%%%%%%%%%%%%%%%%

\section{Gas-Phase Abundances and Dust Depletion Pattern}
\label{sec:depletion}

The column densities of ions measured from absorption lines refer to the species in the gas phase. We assume the ionization correction of metals is negligible due to the shielding effect of large column density of neutral hydrogen from ionizing photons. The total elemental column densities can be derived from the dominant ionization state, i.e., the lowest energy species where the ionization potential is above 13.6 eV. The gas-phase abundances reveal the underlying nucleosynthetic abundances modified by differential depletion pattern of elements. Heavy element abundances are measured relative to solar values, as [X/H] $\equiv$ log $N(X)/N(H)$ - log $(X/H)_{\sun}$ assuming that $N$(H) = $N$(H~{\sc i}). The resultant metal abundances are listed in Table \ref{tab:metal}. The Zn~{\sc ii} column density indicates that this is a metal-strong DLA (log $N$(Zn~{\sc ii}) $\geq$ 13.15) as defined by \cite{HerbertFort06}.

\begin{figure*}
\centering
{\includegraphics[width=13cm, height=9cm]{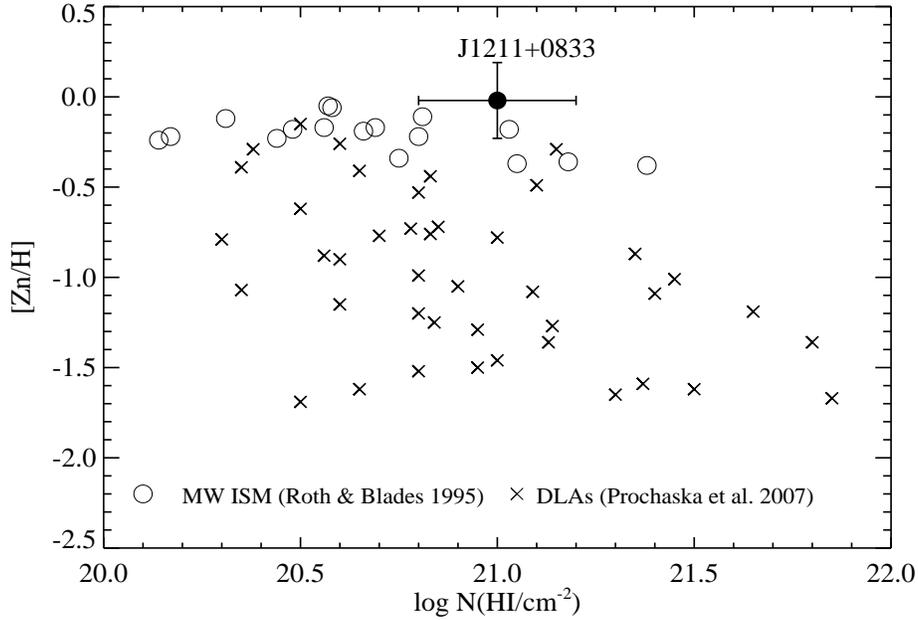}}   
\caption{Metallicity comparison with MW clouds \citep{Roth95} and normal DLAs \citep{Prochaska07} in terms of [Zn/H]. The J1211+0833 absorber, with a moderate hydrogen column density, distinguishes itself by having a metallicity ([Zn/H] = -0.07 $\pm$ 0.21) comparable to that of the MW clouds and above all the DLAs in this sample.}
\label{fig:Zn_H}
\end{figure*}

Fig.\ref{fig:Zn_H} demonstrates metallicity measured in [Zn/H] versus hydrogen column density log $N$(H~{\sc i}) in different absorbers.  The crosses are the DLAs with [Zn/H] measurements in \cite{Prochaska07} and the open circles are clouds in the MW by \cite{Roth95}. The J1211+0833 absorber, with a moderate hydrogen column density (in the DLA regime), distinguishes itself by having a metallicity ([Zn/H] = -0.07 $\pm$ 0.21) comparable to that of the MW clouds and above all the DLAs in this sample.

%%%%%%%%%%%%%%%%%%%%%%%
%metallicity summary table (Keck & UVES )
%%%%%%%%%%%%%%%%%%%%%%%

\begin{table}
\centering
\caption{Gas-phase abundances of the absorber towards J1211+0833 using Keck/ESI and VLT/UVES. }
\begin{tabular}{@{}lccc@{}}
\hline\hline
  & J1211+0833  & source &Solar abundances \\
\hline
[Zn/H] & -0.07$\pm$0.21  & VLT/UVES  & -7.35\\ 
$ $[Si/H] & -0.98$\pm$0.23 & Keck/ESI  &-4.45\\
$ $[Fe/H]& -1.81$\pm$0.21 & Keck/ESI & -4.49\\
$ $[Mn/H] & -1.39$\pm$0.23 & Keck/ESI &-6.47\\
$ $[Ni/H] & -1.63$\pm$0.22 & VLT/UVES &-5.75\\
$ $[Mg/H]& $>$ -2.11   & Keck/ESI &-4.42\\
$ $[Cr/H]& -1.90$\pm$0.24 & VLT/UVES  &-6.32\\
\hline
\end{tabular}
\label{tab:metal}
\end{table}

\begin{figure*}
\centering
{\includegraphics[width=14cm, height=10cm]{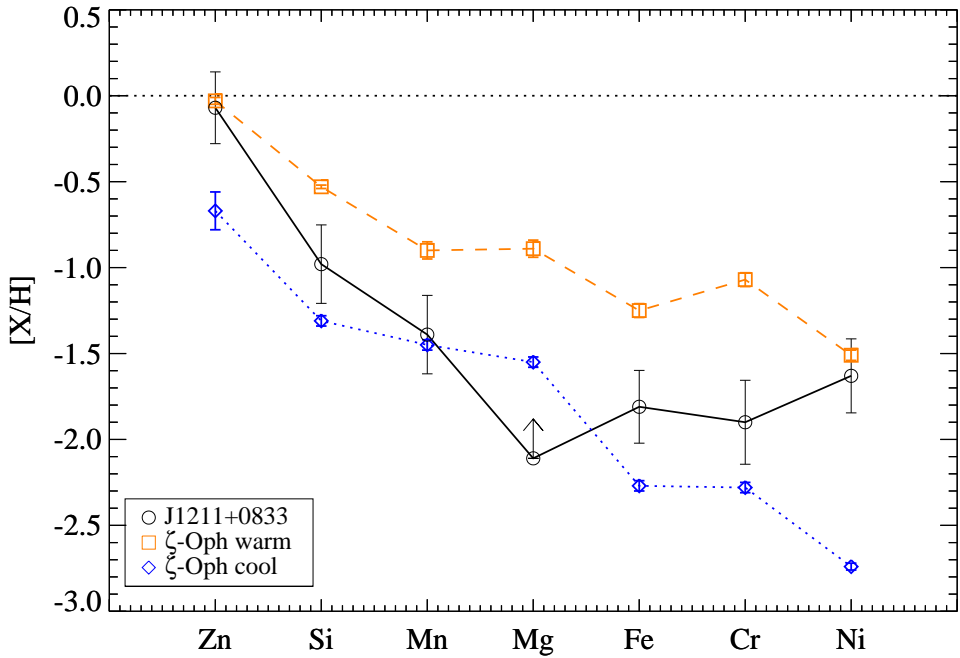}}   
\caption{Dust depletion pattern of the absorber towards J1211+0833 in comparison with that of the warm and cool diffuse disk clouds toward $\zeta$ Ophiuchi \citep{Savage96}. A very high relative abundance of [Zn/Fe] = 1.74 indicates the metals are heavily depleted onto dust grains. The dust depletion level towards J1211+0833 lies between that of the warm disk clouds and the cold disk clouds with an enhanced Zn abundance.  }
\label{fig:depletion}
\end{figure*}

In Fig. \ref{fig:depletion} we show the dust depletion pattern of the absorber towards J1211+0833 compared to that of the MW cool and warm diffuse disk clouds toward $\zeta$ Ophiuchi \citep{Savage96}. A very high relative abundance of [Zn/Fe] = 1.74 indicates the metals are heavily depleted onto dust grains. The dust depletion level of the absorber towards J1211+0833 lies between that of the warm disk clouds and the cold disk clouds with an enhanced Zn abundance. The high depletion level supports the existence of a large number of dust grains and thus the observed strong 2175 \AA$ $ dust extinction bump.

\section{Physical conditions through atomic and molecular lines}
\subsection{C~{\sc i} profiles and excitation temperatures}

\begin{figure*}
\centering
{\includegraphics[width=15cm, height=12cm]{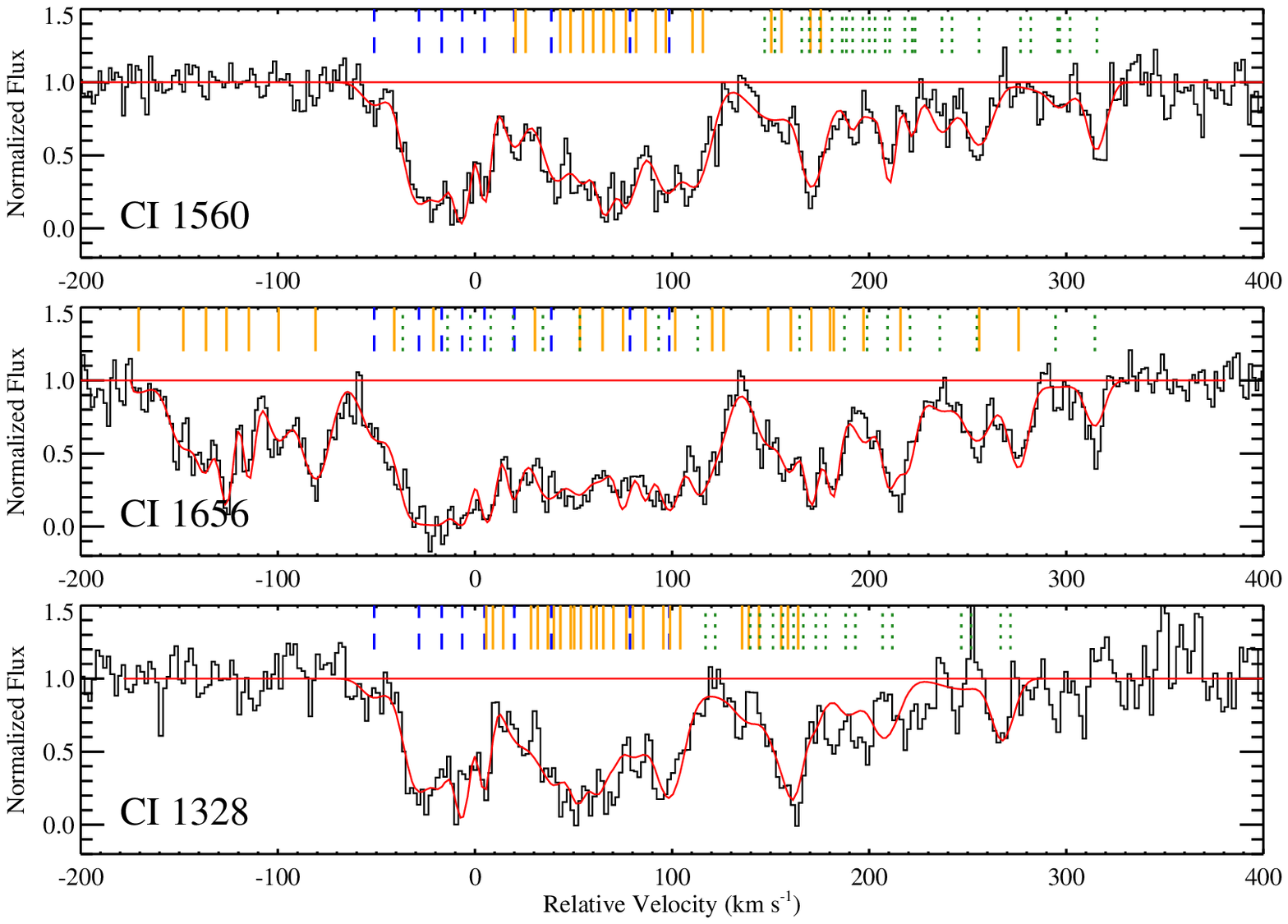}}   
\caption{C~{\sc i} velocity structure (nine components) derived from simultaneous fitting of three multiplets: C~{\sc i} 1560, C~{\sc i} 1656, and C~{\sc i} 1328 in the VLT/UVES data. Black is the data and the overall fit is in red. The different fine structure transitions are color-coded by the vertical lines ----- C~{\sc i}: blue (dashed lines), C~{\sc i}$^*$: orange (solid lines), C~{\sc i}$^{**}$: green (dotted lines)}.
\label{fig:CI}
\end{figure*}

\begin{figure}
{\includegraphics[width=9cm, height=6cm]{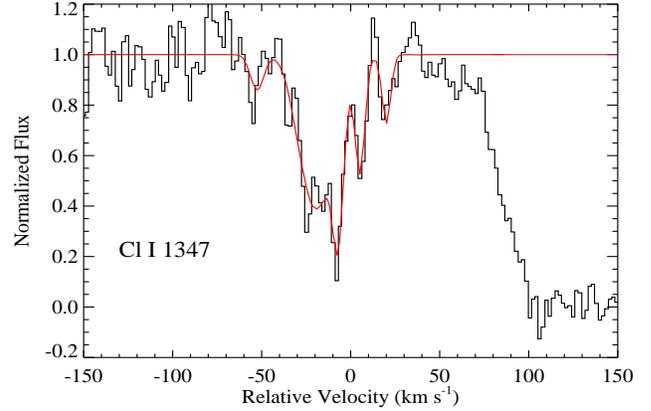}}    
\caption{A multi-component fit on Cl~{\sc i} $\lambda$1347 from VLT/UVES. The resultant total neutral chlorine column density is log $N$(Cl~{\sc i}) = 13.95 $\pm$ 0.10 $\cmsq$. }
\label{fig:chlorine}
\end{figure}

The strongest multiplets at 1560 \AA$ $ and 1656 \AA$ $ are detected in the Keck/ESI spectra, and many close fine structure transitions are heavily blended. We use the VPFIT package to simultaneously fit as many C~{\sc i} multiplets as possible in the UVES data: the strongest multiplets at 1560 \AA$ $ and 1656 \AA$ $ that fall redward of the Ly$\alpha$ forest and the multiplet at 1328 \AA$ $. The other multiplets at bluer wavelengths are rejected due to low signal-to-noise ratio and contamination. Fig. \ref{fig:CI} demonstrates the three multiplets in black, overlaid with our fit in red, and the C~{\sc i} fine structure transitions are marked with the vertical lines. Each fine structure state is denoted by a different color (C~{\sc i}: blue, C~{\sc i}$^*$: orange, C~{\sc i}$^{**}$: green). As shown in the figure, nine velocity components are required to produce a satisfactory fit, which is similar to the velocity structure of low ionization lines obtained from Zn~{\sc ii}, Fe~{\sc ii}, Ni~{\sc ii}, and Cr~{\sc ii}. The components are located at $v$ $\sim$ -51 \kms, -28 \kms, -17 \kms, -6 \kms, +5 \kms, +20 \kms, +39 \kms, +79 \kms, and +99 \kms. A multi-component fit on a neutral chlorine line (Fig. \ref{fig:chlorine}), Cl~{\sc i} $\lambda$1347,  reveals a similar velocity structure with C~{\sc i}, which is expected due to their similar ionization potentials \citep{Jura74} (more discussion on chlorine in Section \ref{sec:Chlorine}).

%%%%%%%%%%%%%%%%%%%
%CI multiplets
%%%%%%%%%%%%%%%%%%%

The resultant C~{\sc i}, C~{\sc i}$^*$, C~{\sc i}$^{**}$ column densities in each component are summarized in Table \ref{tab:NCI}. We also derive the excitation temperature ($T_{ex}$) of C~{\sc i} according to the Boltzmann equation,
\begin{equation}
N_e/N_g = g_e/g_g {\rm exp}(-\Delta E_{eg}/kT_{ex}),
\end{equation}
where $N_e$ and $N_g$ are column densities in the excited and ground state levels, and $g_e$ and $g_g$ are the corresponding statistical weights. k is the Boltzmann constant. $\Delta E_{eg}$ (if given in K) is 23.6 K for the $J$=1 to 0  transition in C~{\sc i}. The excitation temperatures derived from the C~{\sc i} and C~{\sc i}$^*$ states in each component are all consistent with and higher than the cosmic microwave background (CMB) temperature $T_{CMB}$ =  8.5 K at the absorber's redshift. $T = T_0(1+z)$ where $T_0$ = 2.725 K \citep{Mather99}. 

\subsection{Deriving physical conditions through C~{\sc i} fine structure lines and the C~{\sc ii}/C~{\sc i} ratio}

In order to derive the physical conditions in the absorbing gas, we employ the method by \cite{Jorgenson10} which simultaneously constrains both the volume density and temperature of the absorbing gas without making assumptions of temperatures. We invoke the assumptions of steady state and ionization equilibrium. Various excitation and de-excitation mechanisms are taken into account: spontaneous radiative decay, direct excitation by the CMB, UV pumping due to a radiation field, and collisional excitation and de-excitation. The volume densities and temperatures are constrained through two steps: first solving the steady state equation by calculating the ratio of each excited state relative to the ground state, and then further constraining the C~{\sc i} solutions with the ionization equilibrium invoked by the C~{\sc ii}/C~{\sc i} ratio. 

The rate of populating and de-population of state $i$ is given by 

\begin{equation}
\begin{array}{rcl}
\sum_j n_j (A_{ji} + B_{ji} u_{ji} + \Gamma_{ji} + \sum_k n^k q^k_{ji}) = \\
n_i \sum_j (A_{ij} + B_{ij} u_{ij} + \Gamma_{ij} + \sum_k n^k q^k_{ij})
\end{array}
\label{eqn:steadystate}
\end{equation}

\noindent where $A_{ij}$ is the probability of spontaneous decay , $B_{ij}$ is the probability of stimulated emission, $u_{ij}$ is the energy density of the cosmic microwave background radiation field, $\Gamma_{ij}$ is the indirect excitation rate due to fluorescence defined by \cite{silva01}.
In addition, the excitation and de-excitation terms due to collisions are included. The quantity $n^k$ is the volume density of the collision partner $k$, where $k$ can be atomic hydrogen, electron, and proton (densities denoted as $n$(H~{\sc i}), $n_e$, and $n_p$) in the case of a DLA and $q^k_{ij}$ = $< \sigma v>$ is the collision rate coefficient, where $\sigma$ is the cross section and $v$ is the relative speed between the collision partners. The reverse rates are calculated using the assumption of detailed balance. Taking all of these excitation and de-excitation mechanisms into account, the ratios of the upper to lower fine structure level populations of $C^0$ become a function of redshift, temperature, neutral hydrogen volume density, and electron density. We first consider the UV pumping due to a radiation field that has a typical intensity of $J_{\nu}$ $\sim$ $10^{-19}$ \intensity measured in damped Ly-$\alpha$ systems \citep{Wolfe04} and an electron density of $n_e$ = 10$^{-3}$ $n$(H~{\sc i}) which is consistent with the typical values found in DLAs (\citealt{Srianand05,Neeleman15}). For now, we use typical values but note that they could actually differ strongly in this peculiar system. We discuss the sensitivity to the radiation field and electron fraction afterwards.

%%%%%%%%%%%%%%%%%%%
%Cl I 1347
%%%%%%%%%%%%%%%%%%%

\begin{figure*}
\centering
{\includegraphics[width=8.5cm, height=6.5cm]{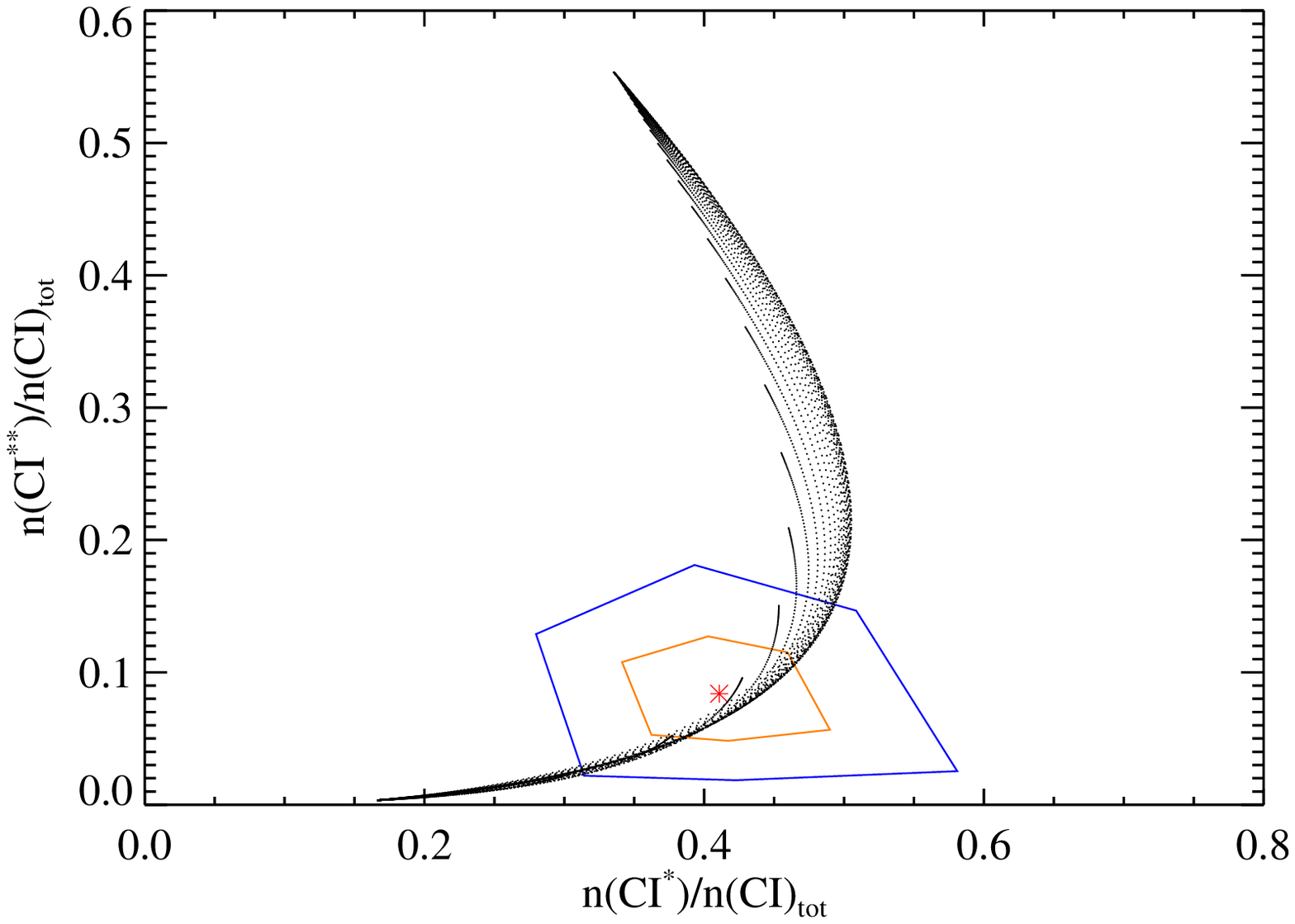}}    
{\includegraphics[width=8.5cm, height=6.5cm]{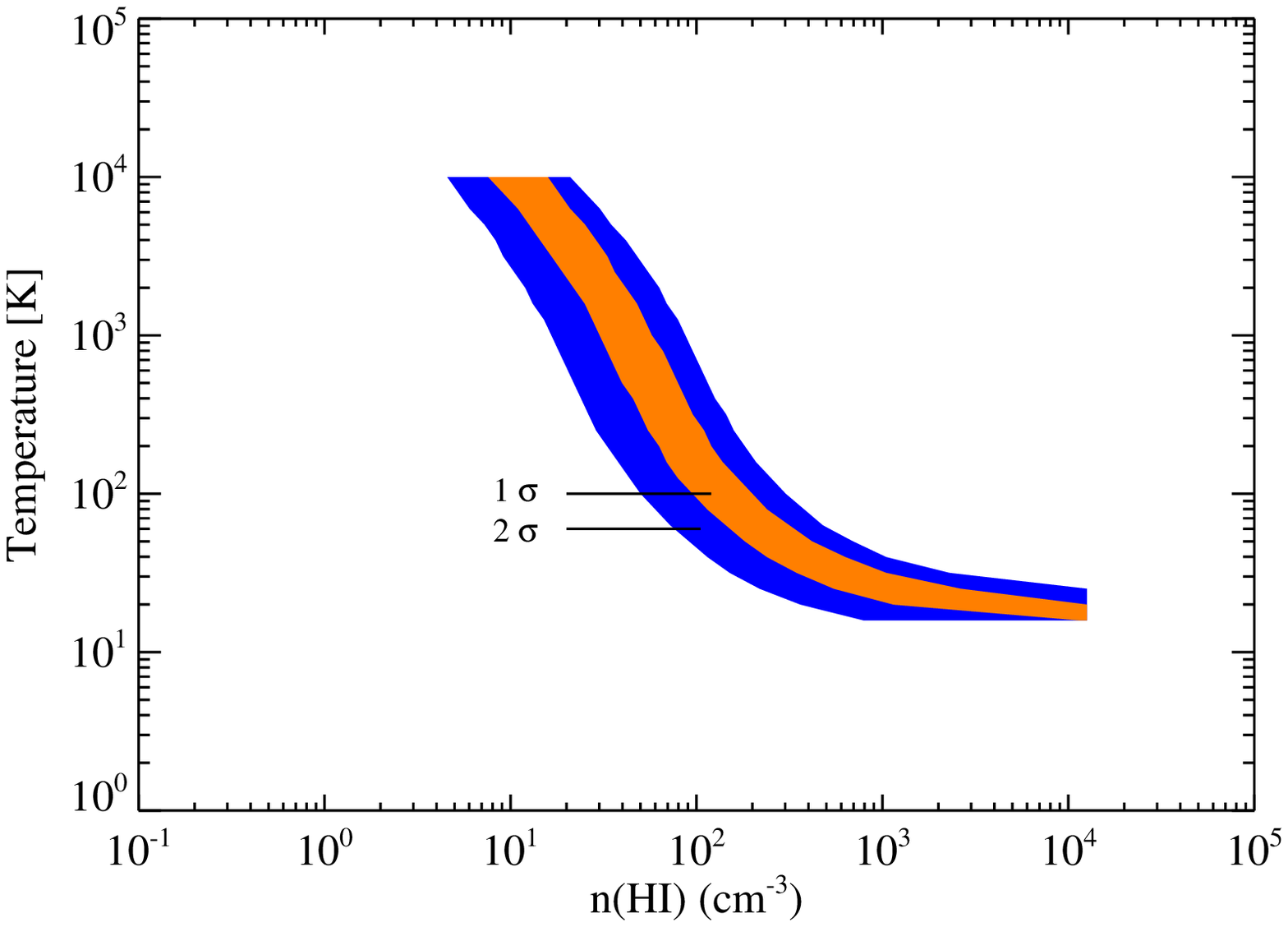}}    
{\includegraphics[width=8.5cm, height=6.5cm]{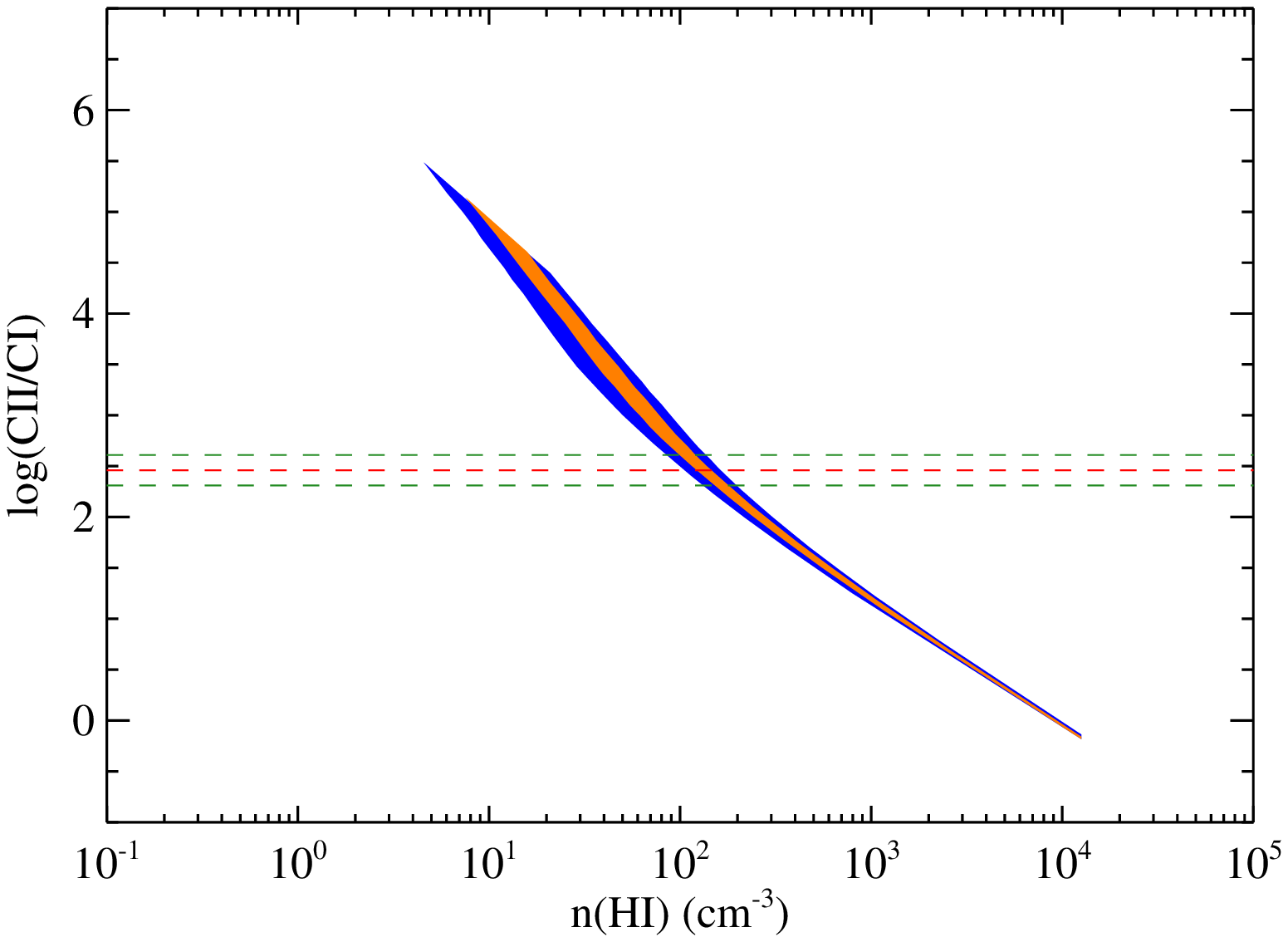}}
{\includegraphics[width=8.5cm, height=6.5cm]{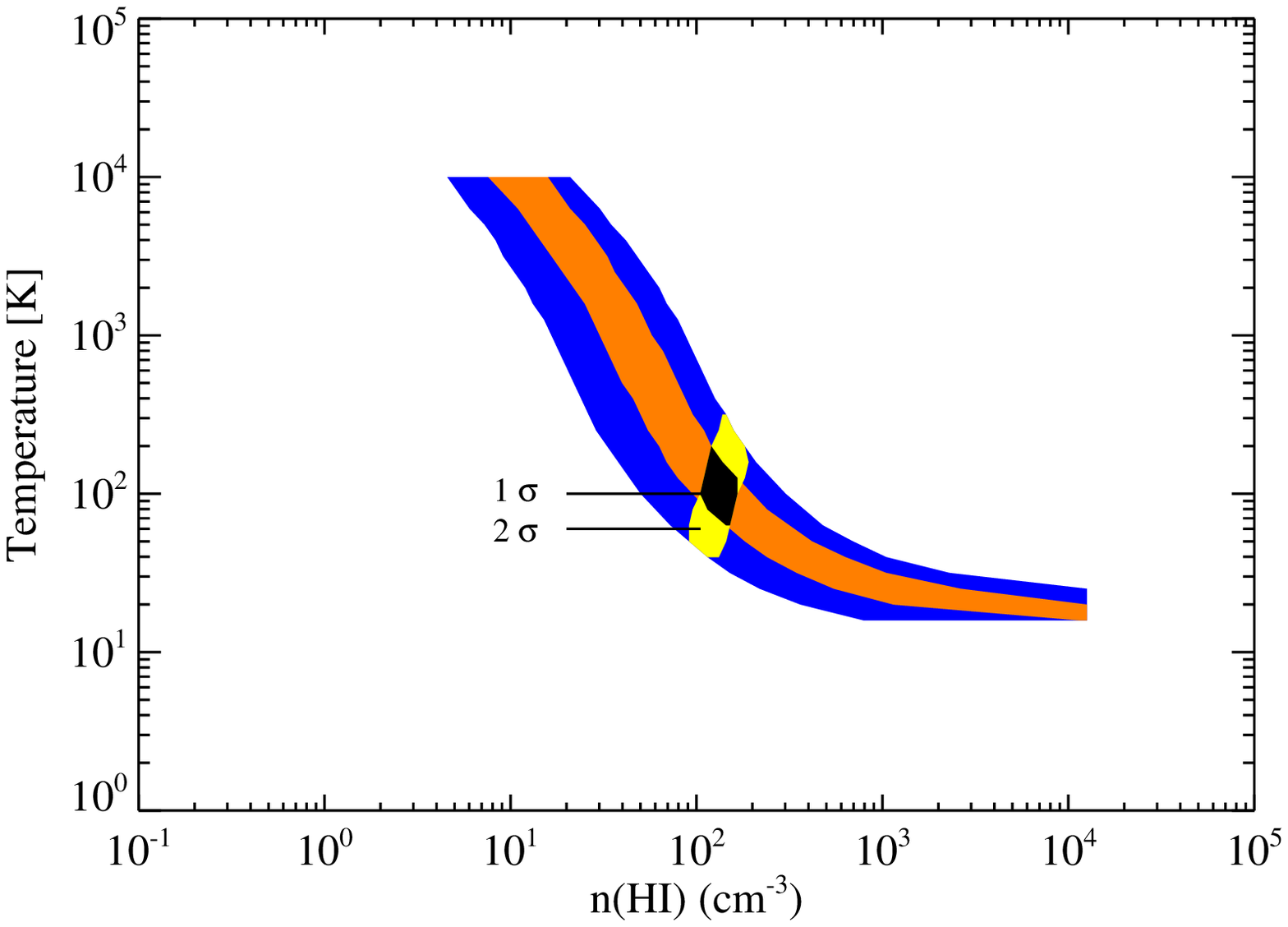}}    
\caption{{\it Top Left}: $f1$ vs. $f2$ for component 3 of the J1211+0833 absorber. The measured value is denoted by a red asterisk and the 1 $\sigma$ (2 $\sigma$) error polygon is in orange (blue). The black curves are the theoretical tracks which cover possible ranges of temperatures and hydrogen volume densities. {\it Top Right}: Temperature vs. $n$(H~{\sc i}) for the allowed solutions (1 $\sigma$ in orange and 2 $\sigma$ in blue) constrained by the C~{\sc i} fine structure lines only. {\it Bottom Left}: log $N$(C~{\sc ii}/C~{\sc i}) vs. n(H~{\sc i}) for the allowed solutions further constrained by introducing ionization equilibrium. The measured log $N$(C~{\sc ii}/C~{\sc i}) is indicated by the red dashed line with the 0.15 dex error in green. {\it Bottom Right}: The final solutions are shown in black (1 $\sigma$) and yellow (2 $\sigma$). }
\label{fig:CI_analysis}
\end{figure*}

We find the solution in the ($f1$, $f2$) plane where $f1$ $\equiv$ $n$(C~{\sc i}$^*$)/$n$(C~{\sc i})$_{tot}$, $f2$ $\equiv$ $n$(C~{\sc i}$^{**}$)/$n$(C~{\sc i})$_{tot}$, and $n$(C~{\sc i})$_{tot}$ = $n$(C~{\sc i}) + $n$(C~{\sc i}$^*$) +$n$(C~{\sc i}$^{**}$). The values of ($f1$, $f2$) for each component are given in Table \ref{tab:NCI}. We show an example case of component 3 of the absorber towards J1211+0833 in Fig. \ref{fig:CI_analysis}. According to the Zn~{\sc ii} column density distribution in each component, component 3 shows the highest column density with about 30\% of the total. We refer to the Zn~{\sc ii} column density not only because its velocity profile is the best measured one available to us but also the fact that Zn is non-refractory and non-depleted such that it is likely to trace the underlying relative distribution of H~{\sc i} or other low ions in each cloud.  We generate theoretical tracks: one for each temperature from T = 10 to 10$^4$ K in steps of 0.1 dex; for each track $n$(H~{\sc i}) ranges from 10$^{-3.5}$ to 10$^{4.1}$ cm$^{-3}$ in steps of 0.02 dex. The orange/blue polygon, which is determined by the 1 $\sigma$/2 $\sigma$ error bars for the column densities, encloses the region where 1 $\sigma$/2 $\sigma$ solutions are accepted (Top Left). $n$(H~{\sc i}) is well constrained for each temperature, but temperature is still a free parameter based on C~{\sc i} fine structure lines only (Top Right). 

We further constrain the C~{\sc i} solutions by introducing the C~{\sc ii}/C~{\sc i} ratio under the assumption of ionization equilibrium. The ionization equilibrium in this case can be written as  $n_e$ $n$(C~{\sc ii}) $\alpha$ = $n$(C~{\sc i}) $\Gamma$, where $\alpha$ is the recombination coefficient of element C$^{+}$ to C$^{0}$ and $\Gamma$ represents the ionization rate which is proportional to the radiation field intensity.  The recombination coefficient $\alpha$ includes the effects of radiative plus dielectric recombination \citep{Shull82} and grain-assisted collisional recombination \citep{Weingartner01}.  We cannot directly measure the $N$(C~{\sc ii}) because the resonance lines of C~{\sc ii} are too saturated.  We instead measure $N$(C~{\sc ii}) by proxy using $N$(Si~{\sc ii}) as in \cite{Wolfe04}: [C/H] = [Si/H] $+$ [Fe/Si]$_{int}$ where the intrinsic (nucleosynthetic) ratio [Fe/Si]$_{int}$ = $-$0.2 for a minimal depletion model or [Fe/Si]$_{int}$ = 0.0 for a maximal depletion model.  We adopt an average [Fe/Si]$_{int}$ ratio of the two depletion models and account for the errors induced by choice of depletion models. We use $N$(H~{\sc i}) and $N$(C~{\sc ii}) as global measurements and apply them to individual C~{\sc i}-bearing clouds. Since Ly-$\alpha$ is damped, the velocity structure associated with each H~{\sc i} cloud is unknown. For C~{\sc ii}, we cannot measure an accurate column density in each velocity component unless an unsaturated C~{\sc ii} or Si~{\sc ii} line is covered but unfortunately Si~{\sc ii} 1808 lies in the gap of the VLT/UVES spectrum. We do not attempt to perform strict individual component analysis in this regard.

%%%%%%%%%%%%%%%%%%
%CI analysis plots
%%%%%%%%%%%%%%%%%%

We plot log(C~{\sc ii}/C~{\sc i}) as a function of $n$(H~{\sc i}) to demonstrate the constraints imposed by ionization equilibrium (shown in the lower left panel of Fig. \ref{fig:CI_analysis}). The measured log $N$(C~{\sc ii}/C~{\sc i}) = 2.46 $\pm$ 0.15 for component 3 of C~{\sc i}-bearing clouds is indicated by the horizontal dashed lines. The final solutions are re-plotted in the temperature vs. $n$(H~{\sc i}) diagram (Bottom Right panel). In the case of component 3,  the density is constrained in the range 105 $\la$ $n$(H~{\sc i}) $\la$ 166 cm$^{-3}$ at the 1 $\sigma$ level and 91 $\la$ $n$(H~{\sc i}) $\la$ 191 cm$^{-3}$ at the 2 $\sigma$ level; the temperature is constrained to be 63 K $\la$ T $\la$ 200 K (1 $\sigma$)  and 40 K $\la$ T $\la$ 316 K (2 $\sigma$). The constrained pressure is 3.66 $\la$ log ($P$/k) $\la$ 4.66 cm$^{-3}$ K (2 $\sigma$). We can compare  our results to those found in the local ISM. Jenkins $\&$ Tripp (2011) find by using C~{\sc i} fine structure lines that the cold neutral medium in the local ISM has an average pressure of log ($P$/k) = 3.58 $\pm$ 0.18 cm$^{-3}$ K.

We further estimate the characteristic size and mass of the absorbing cloud. The size can be estimated by the ratio of $N$(H~{\sc i}) to n(H~{\sc i}): $l$ = $N$(H~{\sc i})/$n$(H~{\sc i}) $\sim$ 2 - 4 pc. This is essentially an upper limit in a cold component because $N$(H~{\sc i}) is distributed over all components. Therefore C~{\sc i} traces very dense pockets of very cold gas at slightly higher pressure. The associated gas cloud has a mass $M$ = $m_pn_{HI}l^3$ ($m_p$ is proton mass) $\sim$ 3110 - 9000 solar masses. A rough estimate of the mass of the galaxy is $M_{gal}$ = $v^2$R/G $\sim$ 5 $\times$ 10$^{10}$ ($\frac{R}{10  \rm{kpc}}$) M$_{\sun}$, where $v$ is the velocity span of the components, R is the radius, and G is the gravitational constant.

%%%%%%%%%%%%%%%%%%%
%CI fine structure lines
%%%%%%%%%%%%%%%%%%%

\begin{table*}
\centering
\caption{VPFIT results for the C~{\sc i} data from VLT/UVES }
\begin{tabular}{cccccccc}
\hline\hline
Comp. & $z_{abs}$   &  $b$       &  log $N$(C~{\sc i})   &  log $N$(C~{\sc i}$^*$)  & log $N$(C~{\sc i}$^{**}$) & $T_{ex}$ & ($f1$,$f2$) \\
            &          & \kms &  $\cmsq$    &  $\cmsq$    & $\cmsq$       & K           &      \\
\hline
1    &2.116069  &7.42$\pm$2.35        &12.88$\pm$0.10  &12.66$\pm$0.25  &12.54$\pm$0.38 &14.6  & (0.29,0.22) \\
2    &2.116305  &8.18$\pm$0.62      &13.94$\pm$0.04  &13.55$\pm$0.05  &12.92$\pm$0.17 &11.8  & (0.27,0.06)\\
3    &2.116425  &4.77$\pm$0.99      &13.63$\pm$0.07  &13.54$\pm$0.05  &12.85$\pm$0.16 &18.0  & (0.41,0.09)\\
4    &2.116533  &2.62$\pm$0.33      &14.55$\pm$0.32  &13.90$\pm$0.06  &13.61$\pm$0.07 &9.1   & (0.17,0.09)\\
5    &2.116651  &2.05$\pm$0.36      &13.96$\pm$0.26  &13.70$\pm$0.07  &13.07$\pm$0.09 &13.9  & (0.33,0.08)\\
6    &2.116808  &7.13$\pm$0.92      &13.33$\pm$0.06  &13.43$\pm$0.05 &13.20$\pm$0.07 &27.0  & (0.42,0.25)\\
7    &2.117003  &8.05$\pm$0.55      &13.43$\pm$0.06  &13.80$\pm$0.03  &13.51$\pm$0.04 &101.4 & (0.52,0.26)\\
8    &2.117418  &15.34$\pm$2.54     &13.53$\pm$0.08  &13.41$\pm$0.08  &12.91$\pm$0.16 &17.2  & (0.38,0.12)\\ 
9    &2.117625  &5.38$\pm$0.47      &13.48$\pm$0.07  &13.66$\pm$0.04  &13.40$\pm$0.04 &35.7  & (0.45,0.26) \\
\hline
total &                &                              &14.84$\pm$0.17  &14.56$\pm$0.02  &14.18$\pm$0.03  & 13.5  & (0.30,0.16) \\  
\hline
\end{tabular}
\label{tab:NCI}
\end{table*}

\begin{figure}
\centering
{\includegraphics[width=8.5cm, height=6cm]{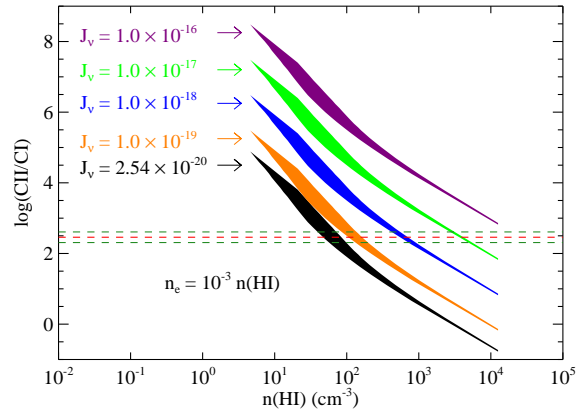}}   
\caption{The sensitivity of C~{\sc ii}/C~{\sc i} ratio to the radiation field intensity in the example case of component 3. Theoretical tracks corresponding to each $J_{\nu}$ are marked by different colors. The electron fraction is held fixed.}
\label{fig:Jv_variation}
\end{figure}

\begin{figure}
\centering
{\includegraphics[width=8.5cm, height=6cm]{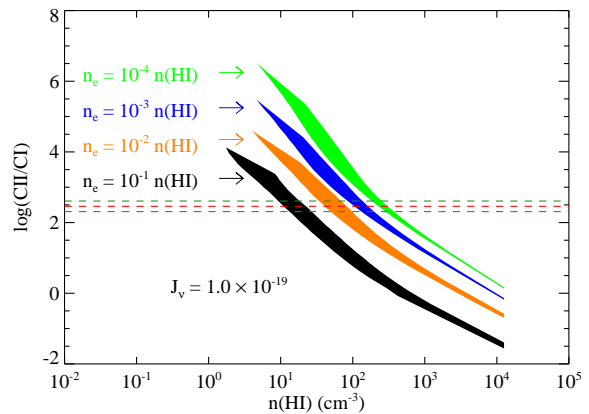}}  
\caption{The sensitivity of C~{\sc ii}/C~{\sc i} ratio to the electron fraction in the example case of component 3. Theoretical tracks corresponding to each n$_e$ are marked by different colors. The radiation field intensity is held fixed.}
\label{fig:ne_variation}
\end{figure}

The above results are based on the typical values of the radiation field intensity $J_{\nu}$ and electron fraction in DLAs. At those high hydrogen volume densities, the relative populations of C~{\sc i} fine structure states are dominated by collisional excitation and de-excitation and the UV pumping has negligible effect. Whereas the C~{\sc ii}/C~{\sc i} ratio can be largely driven by the UV pumping rate as well as the electron fraction as C~{\sc ii}/C~{\sc i} is directly proportional to  $\Gamma$/$n_e$.  We demonstrate how the theoretical C~{\sc ii}/C~{\sc i} ratio moves in the log $N$(C~{\sc ii}/C~{\sc i}) vs. $n$(H~{\sc i}) plane (Fig. \ref{fig:Jv_variation}) with a grid of $J_{\nu}$ from the strength of the Haardt-Madau background (2.54 $\times$ 10$^{-20}$; \citealt{Haardt96}) to 10$^{-16}$ \intensity indicated by different colors. The electron fraction is fixed to $n_e$ = 10$^{-3}$ $n$(H~{\sc i}).  All the tracks are the 2 $\sigma$ results. Again, the allowed solutions are constrained by the measured C~{\sc ii}/C~{\sc i} ratio. The case of $J_{\nu}$ = 10$^{-16}$ is ruled out by the measured value. The allowed solutions are constrained in the range 40 -- 5200 cm$^{-3}$. Fig. \ref{fig:ne_variation} shows the variation due to a grid of possible electron fractions (10$^{-4}$ to 10$^{-1}$) with a fixed $J_{\nu}$ of 10$^{-19}$ \intensity.  The range of $n$(HI) is restricted to be 10 -- 360 cm$^{-3}$.  Only by having good knowledge of these two quantities can we tightly constrain the temperature and density.

The other uncertainty is induced by applying the total $N$(C~{\sc ii}) to individual C~{\sc i}-bearing clouds. Only a part of the total $N$(C~{\sc ii}) is actually associated with each C~{\sc i} cloud. The adopted C~{\sc ii}/C~{\sc i} ratio serves as an upper limit. A decrease in the amount of $N$(C~{\sc ii}) associated with the C~{\sc i} cloud would result in a higher volume density and lower temperature and therefore an approximately constant pressure.

\subsection{Cloudy photoionization modeling}

The physical conditions derived from the above method are average properties of the absorbing cloud or the same condition throughout the cloud. This approach is sensitive to the adopted electron density and radiation field intensity which can be poorly constrained. We further construct a Cloudy (\citealt{Ferland13}; version 13.03) photoionization model of this absorber by which we can utilize observed quantities in addition to C~{\sc i} fine structure lines.  We model the absorber as plane-parallel slabs of gas of constant density exposed to a radiation field. We consider the spectral energy distribution of the Milky Way ISM, table ISM built in Cloudy.  The intensity of the ISM radiation can be changed by a scale factor. We also include the cosmic ray background which is known to influence the ionization and chemical state of high-density ISM. The cosmic microwave background at the absorber's redshift is also taken into account. The chemical abundance is initially set to that of the local ISM but the gas phase and grain abundances can be increased or decreased by a scale factor. We run a grid of simulations for different values of hydrogen volume density $n$(H~{\sc i}) (equivalent to varying the ionization parameter) and subsequently vary the above-mentioned parameters to match the observed column densities of Zn~{\sc ii} and Fe~{\sc ii} in each corresponding velocity component since these two ions have the most reliable individual-component measurements from the available VLT/UVES data. However, these ions are not sensitive to $n$(H~{\sc i}). We include neutral chlorine and magnesium in the modeling of which the column densities are quite sensitive to the hydrogen density. The best model and the best-fit $n$(H~{\sc i}) are determined by $N$(C~{\sc i}) in combination with those ions.  

\begin{figure}
{\includegraphics[width=9cm, height=6.2cm]{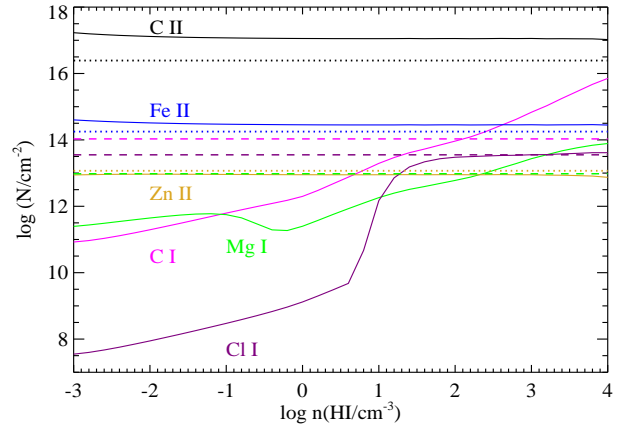}}   
\caption{The Cloudy photoionization model for component 3 shown in the column densities of atoms/ions vs. n(H~{\sc i}) diagram. The measured values are marked with the dashed and dotted lines while the predicted values from Cloudy are the solid lines with the corresponding colors. The metal abundances are scaled to half of the MW ISM. The incident radiation field has an intensity 10 times that of the MW ISM radiation.  }
\label{fig:cloudy}
\end{figure}

Fig. \ref{fig:cloudy} shows the Cloudy model for component 3.  The preferred model is selected not only because the predicted $N$(Zn~{\sc ii}) and $N$(Fe~{\sc ii}) match the measured values but also because Mg I, C I, and Cl I agree on n(HI). The models that do not satisfy these criteria are therefore rejected. As a result, the preferred model gives a metallicity half of that of the MW ISM and a radiation field 10 times the intensity of the MW ISM radiation.  As is often the case, the modeling predicts too much C~{\sc ii} \citep{Fox14}. The predicted neutral carbon column density increases as $n$(H~{\sc i}) rises.  The best-fit $n$(H~{\sc i}) value can be determined as the intercept of the solid (predicted) and dashed (measured) lines: log $n$(H~{\sc i}) $\sim$ 2. The predicted Cl~{\sc i} does not match the observed column density (Sec. \ref{sec:Chlorine}) until log $n$(H~{\sc i}) is greater than $\sim$ 1.5. Mg~{\sc i} indicates a hydrogen volume density of log $n$(H~{\sc i}) = 2.3. They all agree on a high density ($n$(H~{\sc i}) $\sim$ 100 cm$^{-3}$) neutral medium. The photoionization model with this $n$(H~{\sc i}) value indicates that the electron temperature T$_e$ $\sim$ 70 - 110 K. The average values throughout the slabs are consistent with those derived from the C~{\sc i} technique, which is expected.

\subsection{Neutral Chlorine and molecular hydrogen}
\label{sec:Chlorine}

We detect a strong Cl~{\sc i} $\lambda$1347 line with seven velocity components (Fig. \ref{fig:chlorine}) from which we derive a total column density of log $N$(Cl~{\sc i}) = 13.95 $\pm$ 0.10 $\cmsq$. We disregard the region beyond + 50 \kms due to the severe contamination. Neutral chlorine and molecular hydrogen are known to have a tight relation in the cold phase of the local ISM (\citealt{Jura74,Moomey12}). In the presence of H$_2$, singly ionized chlorine converts rapidly through the exothermic reaction Cl$^{+}$ + H$_2$ ---$>$ HCl$^{+}$. Neutral chlorine can be quickly released from several processes especially collisions between HCl$^{+}$ and H$_2$ or electrons.  \cite{Balashev15} studies this relation at high redshift and finds a $\sim$ 5$\sigma$ correlation between $N$(Cl~{\sc i}) and $N$(H$_2$): $N$(Cl~{\sc i}) $\approx$ 1.5 $\times$ 10$^{-6}$ $\times$ $N$(H$_2$), which confirms that neutral chlorine is an excellent tracer of molecule-rich gas. We do not directly detect H$_2$ due to a very low S/N but it essentially exists. Using that relation, we estimate a molecular hydrogen column density of log $N$(H$_2$) $\approx$ 19.74 $\cmsq$ and a molecular fraction of 0.1. Higher S/N spectra would be required to show the H$_2$ absorption lines.

\subsection{Carbon monoxide}

\begin{figure}
{\includegraphics[width=8.5cm, height=14cm]{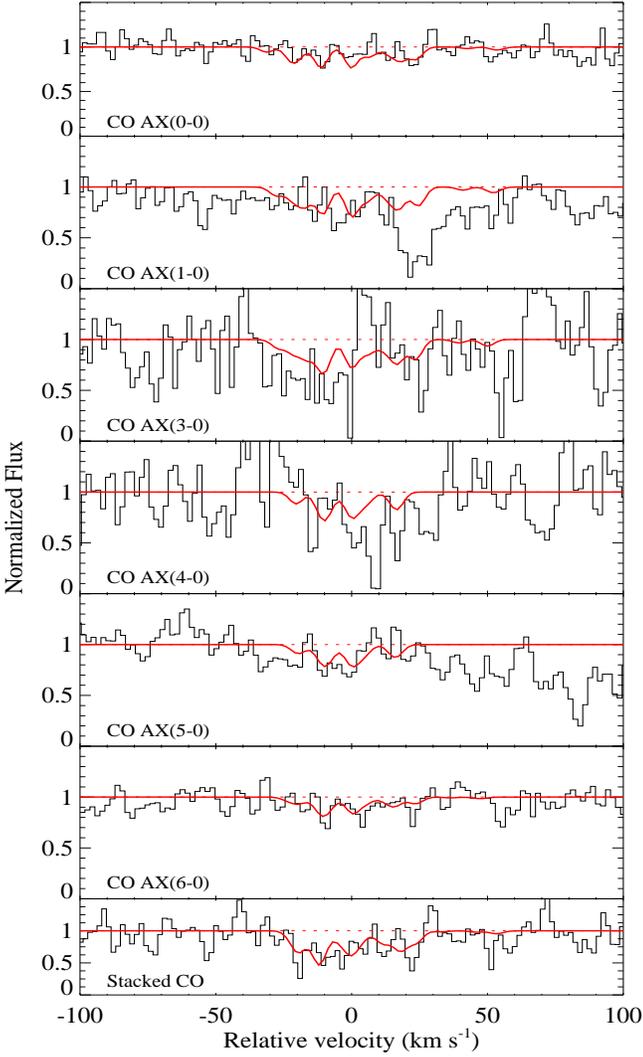}}
\caption{The CO A-X bands with a two-component fit (i.e., component 3 and 4). The rotational levels of $J$ = 0,1,2 are included in the simultaneous fitting. The last panel shows the stacked CO spectrum.}
\label{fig:vpfitCO}
\end{figure}

\begin{figure}
{\includegraphics[width=7.5cm, height=7.5cm]{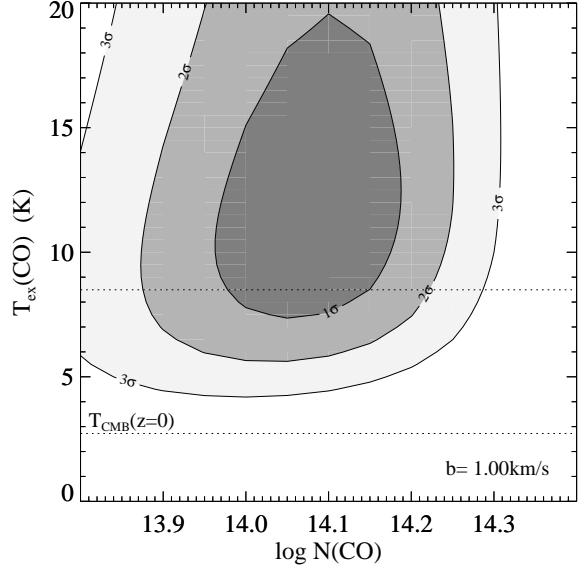}}
\caption{Confidence contours in the T$_{ex}^{CO}$ vs. log $N$(CO) plane assuming $b$ = 1 \kms. The CO column density is log $N$(CO) = 14.1 $\pm$ 0.1 $\cmsq$ with an excitation temperature of $\sim$ 12 K. }
\label{fig:contoursCO}
\end{figure}

The VLT/UVES spectrum covers a series of CO A-X bands. We have simultaneously fitted 6 CO bands, 0-0, 1-0, 3-0, 4-0, 5-0, and 6-0, with a two-component (i.e., velocity component 3 and 4) model. The rotational levels of J = 0, 1, 2 are taken into account. As shown in Fig. \ref{fig:vpfitCO}, individual CO bands are very weak but the feature is detected in the stacked spectrum. This is the second detection with the simultaneous presence of a 2175 \AA$ $ bump beyond the redshift of 2 along a quasar line of sight.  The total CO column density is log $N$(CO) = 14.1 $\pm$ 0.1 $\cmsq$ with an excitation temperature of $\sim$ 12 K. (Fig. \ref{fig:contoursCO}). While the presence of CO is revealed thanks to several coincident absorption lines at the expected position of the electronic bands, and highlighted by stacking the lines, we caution that further observations are required to derive  the parameters accurately, in particular the excitation temperature.

The simultaneous presence of CO and the 2175 \AA$ $ bump indicates that an efficient shielding of the UV radiation field and high metal abundances may be necessary for small carbon-rich grains to exist. These are the ideal conditions for forming molecules efficiently and in particular CO \citep{Sloan08}.

\section{Summary}
\label{sec:summary}

We have reported the detection of a strong Milky Way-type 2175 \AA$ $ dust absorber at $z$ = 2.1166 towards quasar J1211+0833 from the BOSS DR10. Rich metal absorption lines are detected in the Keck/ESI spectrum and the VLT/UVES data reveal a complex velocity structure. It is a remarkable system that has simultaneous detections of C~{\sc i} and CO at the redshift of the 2175 \AA$ $ bump, which are utilized to derive the physical conditions in the absorbing gas. Using the CI fine structure lines and the C~{\sc ii}/C~{\sc i} ratio, we provide constraints on the neutral hydrogen volume density, temperature, and hence pressure.  The results of the analysis are:

1. This 2175 \AA$ $ bump has a bump strength typical of a Milky-way-type bump but is wider than the majority of MW bumps. The distribution of MW bump widths is non-Gaussian with a mean peak and a tail towards larger values, which might suggest two populations of bumps \citep{Fitzpatrick07}. The bump in J1211+0833 belongs to the population in the tail. If PAHs are the carriers, the variation may be interpreted as different mix of PAH molecules \citep{Draine03}.

2. We measure column densities for a large number of metal lines (e.g., Zn~{\sc ii}, Fe~{\sc ii}, Mg~{\sc ii}, Si~{\sc ii}, Ni~{\sc ii}, Mg~{\sc i}, Al~{\sc iii}) covered in Keck/ESI and VLT/UVES. We obtain a hydrogen column density of log $N$(H~{\sc i}) = 21.00 $\pm$ 0.20 $\cmsq$ from fitting the damped Ly-$\alpha$ profile. With $N$(H~{\sc i}), we derive the absolute chemical abundances and the dust depletion pattern which suggest that this absorption system has a solar Zn abundance and is highly dust depleted with a depletion pattern resembling that of the Milky Way disk clouds.

3. The detailed velocity structure is revealed by the high-resolution VLT/UVES data: a minimum of nine velocity components are required to optimally fit the data and they span a velocity interval of $\sim$ 150 \kms. The profiles exhibit an edge-leading asymmetry with the strongest feature in the blue edge. The asymmetry shown in the profile is consistent with the prediction by a simple model of a rotating disk \citep{Prochaska97}.

4. Higher resolution observations with the VLT/UVES also reveal the presence of neutral carbon and CO molecules in the absorbing gas. The C~{\sc i} fine structure line analysis suggests that the physical conditions in the absorbing gas are consistent with  a canonical (T $\sim$ 100 K) cold neutral medium. C~{\sc i} traces dense pockets of cold gas at slightly higher pressures. We have further constructed a Cloudy photoionization model which utilizes atoms/ions other than C~{\sc i} to constrain the physical conditions in the absorbing cloud. Neutral carbon, chlorine, and magnesium mutually agree on a hydrogen volume density of $n$(H~{\sc i}) $\sim$ 100 cm$^{-3}$. 

5. It has been suggested (e.g., \citealt{Eliasdottir09,Noterdaeme09,Prochaska09}) that a correlation exists between the presence of the 2175 \AA$ $ bump feature and prominent C~{\sc i}. The remarkable aspect of this quasar absorber is that carbon monoxide is simultaneously present along with C~{\sc i} and the 2175 \AA$ $ bump. Since the carriers of the 2175 \AA$ $ are generally believed to be carbonaceous material that requires neutral carbon and molecules for grains to form and grow \citep{Henning98}, the simultaneous presence of C~{\sc i}, CO, and the 2175 \AA$ $ bump would not be surprising but rather preferred. Combined with the high metallicity, high dust depletion level, and overall low ionization state of the gas,  all the evidence supports the scenario in which the presence of the bump requires an evolved stellar population (i.e., AGB stars) (\citealt{Noll07,Eliasdottir09}).  The host of the 2175 \AA$ $ bump towards J1211+0833 is likely to be an evolved and chemically-enriched disk galaxy.

\section*{Acknowledgments}

We are very grateful to the anonymous referee for detailed comments on the manuscript. We thank C\'edric Ledoux for help in preparing the VLT/UVES observations and thank Fred Hamann for helpful discussions on the Cloudy model.  We also thank Marcel Neeleman for help in deriving physical conditions through alternative approach. PC is grateful to the Institut d'Astrophysique de Paris for hospitality during the time part of this work was done and to the Ecole Normale Sup\'erieure for financing his internship. 

This work has made use of data obtained by the SDSS-III, Keck, and VLT.  Funding for SDSS-III has been provided by the Alfred P. Sloan Foundation, the Participating Institutions, the National Science Foundation, and the U.S. Department of Energy Office of Science. The SDSS-III web site is http://www.sdss3.org/. SDSS-III is managed by the Astrophysical Research Consortium for the Participating Institutions of the SDSS-III Collaboration including the University of Arizona, the Brazilian Participation Group, Brookhaven National Laboratory, Carnegie Mellon University, University of Florida, the French Participation Group, the German Participation Group, Harvard University, the Instituto de Astrofisica de Canarias, the Michigan State/Notre Dame/JINA Participation Group, Johns Hopkins University, Lawrence Berkeley National Laboratory, Max Planck Institute for Astrophysics, Max Planck Institute for Extraterrestrial Physics, New Mexico State University, New York University, Ohio State University, Pennsylvania State University, University of Portsmouth, Princeton University, the Spanish Participation Group, University of Tokyo, University of Utah, Vanderbilt University, University of Virginia, University of Washington, and Yale University. The W.M. Keck Observatory is operated as a scientific partnership among the California Institute of Technology, the University of California and the National Aeronautics and Space Administration. The Observatory was made possible by the generous financial support of the W.M. Keck Foundation.

\bibliographystyle{mn2e}
\bibliography{J1211}
\label{lastpage}

\end{document}